\documentclass{article}

\usepackage[utf8]{inputenc}
 
\title{Sections and Chapters}
\author{Gubert Farnsworth}
\date{ }
\usepackage{arxiv}
\usepackage{cite}
\usepackage{amsmath,amssymb,amsfonts}
\usepackage{algorithmic}
\usepackage{graphicx}
\usepackage{textcomp}
\usepackage{xcolor}
\usepackage{tabulary}
\usepackage{csquotes}
\newtheorem{defi}{\textbf{Definition}}
\newtheorem{remark}{Remark}
\newtheorem{example}{Example}

\newtheorem{assume}{\textbf{Assumption}}
\newtheorem{lemma}{Lemma}
\usepackage{framed}
\usepackage{nomencl} 
\usepackage{adjustbox}
\usepackage[utf8]{inputenc} 
\usepackage[T1]{fontenc}    
\usepackage{hyperref}       
\usepackage{url}            
\usepackage{booktabs}       
\usepackage{amsfonts}       
\usepackage{nicefrac}       
\usepackage{microtype}      
\usepackage{lipsum}
\usepackage{graphicx}
\usepackage{accents}
\newtheorem{theorem}{Theorem}
\newtheorem{prop}{Proposition}
\graphicspath{ {./images/} }
\newcommand*{\dt}[1]{%
   \accentset{\mbox{\large\bfseries .}}{#1}}

\def\BibTeX{{\rm B\kern-.05em{\sc i\kern-.025em b}\kern-.08em
    T\kern-.1667em\lower.7ex\hbox{E}\kern-.125emX}}

\usepackage{xcolor}
\def\BibTeX{{\rm B\kern-.05em{\sc i\kern-.025em b}\kern-.08em
    T\kern-.1667em\lower.7ex\hbox{E}\kern-.125emX}}

\title{
Passivity and Immersion based-Modified Gradient Estimator: A Control Perspective in Parameter Estimation 
}

\author{
Syed Shadab Nayyer, G. Revati, S. R. Wagh, and N. M. Singh\\
  Control and Decision Research Center (CDRC), EED,\\ Veermata Jijabai Technological Institute (VJTI), Mumbai, India\\ 
  \texttt{sisyed\_p19@ee.vjti.ac.in} \\
}

\begin{document}
\maketitle
\begin{abstract}
In this paper, a constructive and systematic strategy with more apparent degrees of freedom to achieve the accurate estimation of unknown parameters via a control perspective is proposed. By adding a virtual control in the final equation of the gradient dynamics,  the Gradient Estimator (GE) and Memory Regressor and Extension (MRE) approaches are extended. The solution of the virtual control law is identified by the P\&I approach \cite{ShadabArxiv}. The P\&I approach \cite{ShadabArxiv} is based on the choice of an appropriate implicit manifold and the generation of a suitable passive output and a related storage function. This facilitates the virtual control law being obtained in a way that the parametric error converges asymptotically to zero.  Because the above ideas connect with the P\&I approach and GE, the developed methodology is labeled the “passivity and immersion-based modified gradient estimator (MGE)”. The proposed P\&I-based modified gradient estimator is extended via the MRE approach. This modification provides improved transient response and fast convergence. Based on certain PE and non-PE examples, a comparative analysis is carried out to show the efficacy of the proposed approaches.
\end{abstract}

\tableofcontents
\textbf{Remark: This paper is a rough version and is mainly based on certain examples. As the authors are working on the modifications in this paper, the theorem proof to support the proposed theory will be updated in the next version. }
\section{Introduction}  In many Closed-loop Systems (CLS), the PE condition on the regressor vector is hard to satisfy since it generally relies on the future behavior of the regressor vector and requires that the signal have sufficient enough energy for the entire time span \cite{ShadabDREM}. Additionally, parameter convergence necessitates the PE condition on the regressor (function of system states), which depends on parameter convergence and ends up creating a circular argument \cite{nguyen2018model}. The extremely weak and online-verifiable Initial Excitation \cite{roy2019relaxing}-or-Interval Excitation (IE) condition \cite{chowdhary2010concurrent} requires the excitation of a regressor vector in a finite time interval. Latest developments in data-driven and learning methodologies have shown encouraging and promising results in parameter convergence and output tracking through IE conditions. The first parameter estimator that offers reliable and well-established results under IE conditions is the concurrent \cite{chowdhary2010concurrent} and composite learning approaches.

 A dynamic data stack is built to record and store online data discretely, sufficiently rich data is recorded for satisfying IE. Concurrent Learning (CL) makes use of both stored/recorded data and instantaneous data for exponential parameter convergence. However, the derivation of all the system states is required for calculating prediction errors which may not be always feasible. Composite learning \cite{pan2015composite} is another way to estimate parameters based on data storage. It is utilized to retain the exponential convergence property and, to a certain extent, avoids the shortcomings of CL.  Contrary to CL, a second-order filter is implemented in composite learning to address the system state derivative problem. One of the parameters is the integration window length, which is not always calculable in advance. The proposed IE-based parameter estimation results are developed to address the issues with composite learning in \cite{roy2019relaxing}. These approaches don't need to know the window length because they verify the determinant online to determine it.

Regressor filtration gives the opportunity to implement the following principle: the higher the adaptation rate, the higher the parameter convergence rate (and, as a
result, the approximation quality). If filtration is not used, then there is an optimal value of
the adaptation rate. If we exceed it, then we face a convergence rate decrease. Filtration is
also a tool for the regressor extension, for which two main approaches should be mentioned:
The Dynamic Regressor Extension (DRE) \cite{aranovskiy2016parameters, ortega2020new} and the Memory Regressor Extension (MRE) \cite{kreiselmeir, ortega2020new}. In the case of the DRE, many different filters  are applied
to the regressor vector  and the output function. This allows obtaining as many regression equations as the number of the unknown regression parameters. At the same time, it is necessary to somehow find parameter values for each filter, and the identification algorithm will start functioning only when the last filter (delay) output becomes a non-zero value. There are different methods proposed in the literation based on the applicability of filters.

The issues associated with CL and composite learning can be tackled with an online estimator that uses Dynamic Regressor Extension and Mixing (DREM) \cite{aranovskiy2016performance, ortega2020new} in which the linear stable filter is applied to linearly parameterized system equation. The DREM is distinct from classical parameter estimators in the following two ways: First, it guarantees that each parameter error is monotonic, which is stronger than the vector norm monotonicity; Second, it promises convergence of parameters by imposing a  non-square integrability condition on the determinant of an extended regression matrix rather than the stringent PE condition \cite{pan2019learning, ShadabDREM}. The authors in \cite{roy2019relaxing} use a simple integrator instead of the LTI filter. Utilization of such an integrator provides second-order dynamics for parameter estimation without any damping. This method is further extended in \cite{glushchenko2021drem, glushchenko2021robust} by replacing an integrator with a new filter $\frac{e^{-\beta t}}{s}$.

In this paper, we have extended the GE and MRE approach by externally adding the virtual control law in the parametric error dynamics. This modification provides improved transient response and fast convergence. The proposed method is discussed in Section 4 with certain PE and non-PE examples. Based on these examples, a comparative analysis is carried out to show the efficacy of the proposed approaches.

\section{Preliminaries and Background materials}
A Linear Regression Equation (LRE) can be used to represent and characterize the system dynamics in a variety of domains, including adaptive pole placement, system identification and filtering \cite{nguyen2018model, ioannou2012robust}, reinforcement learning, and model-reference adaptive control \cite{chowdhary2010concurrent, pan2015composite, ShadabDREM, hozefaparanonpara}, for the stabilization, control, and monitoring of real-world nonlinear systems \cite{shadab2022gaussian, shadab2023finite}.
The underlying structure of any system dynamics can be made a problem of online estimation of unknown parameter vector appearing linearly with regressor vector as:
\begin{equation}\label{1}
    \mathrm{g}= \begin{bmatrix}
\omega_1(\mathrm{t}) & \omega_2(\mathrm{t}) & ..&.. & \omega_{\mathrm{q}}(\mathrm{t})
\end{bmatrix} \begin{bmatrix}
\vartheta_1 \\ \vartheta_2 \\..\\..\\ \vartheta_{\mathrm{q}}
\end{bmatrix}=\mathrm{\omega^T(t)}\hspace{0.075cm}\vartheta\Rightarrow\mathrm{g}=\mathrm{\omega^T(t)}\hspace{0.075cm}\vartheta\
\end{equation}
where $\mathrm{g}:\mathbb{R}^{\mathrm{q}}\rightarrow \mathbb{R}$ represents the output form of a given system model, $\mathrm{\omega(t)}\in \mathbb{R}^{\mathrm{q}}$ is known regressor vector (linear or nonlinear function of system states), and $\vartheta \in\mathbb{R}^{\mathrm{q}} $ are actual parameters appearing linearly. 
\begin{assume}
There is measurement access to the system states $\mathrm{x(t)}=\left [ \mathrm{x_1(t)}, \mathrm{x_2(t)},...,\mathrm{x_n(t)} \right ]^{\mathrm{T}}$. 
\end{assume}
Suppose, $\hat{\vartheta}\in \mathbb{R}^{\mathrm{q}}$ is online identifier of ideal coefficients $\vartheta$; then the mapping $\hat{\mathrm{g}}:\mathbb{R}^{\mathrm{q}}\rightarrow \mathbb{R}$ represents the estimates of $\mathrm{g}$ and given by
\begin{equation}\label{2}
   \mathrm{\hat{\mathrm{g}}(t)}= \mathrm{\omega^T(t)}\hspace{0.075cm}\hat{\vartheta}
\end{equation}
With $\widetilde{\vartheta}=\vartheta-\hat{\vartheta}$ and $\mathrm{\widetilde{\mathrm{g}}(t)}=\mathrm{g(t)}-\mathrm{\hat{\mathrm{g}}(t)}$, the difference between (\ref{1}) and (\ref{2}) provides the error equation 
\begin{equation}\label{3}
   \mathrm{\widetilde{\mathrm{g}}(t)}= \mathrm{\omega^T(t)}\hspace{0.075cm}\widetilde{\vartheta}.
\end{equation}
based on (\ref{2}) and (\ref{3}), it is evident that when $\hat{\vartheta}\rightarrow \vartheta$, it provides $\mathrm{\widetilde{\mathrm{g}}(t)}\rightarrow 0$ as $\mathrm{t}\rightarrow\infty$.  Therefore, the primary objective is to execute an inverse problem \cite{aranovskiy2016parameters, ortega2020new} utilizing output $\mathrm{g}$ and regressors $\mathrm{\omega(t)}$, and to design a parameter update law.

\subsection{Gradient Estimator}
A well-established choice for $\dt{\hat{\vartheta}}$ is the gradient estimator (i.e., gradient-based adaptive law) which updates the adaptive parameters (coefficients) in the direction of maximum reduction of the specified quadratic error function: 
\begin{equation}\label{4}
    \dt{\hat{\vartheta}}=\tau \mathrm{I_{q \times q}} \hspace{0.1cm}\omega\mathrm{(t)}(\mathrm{g}(t)- \omega^{\mathrm{T}}\hat{\vartheta})=\tau \mathrm{I_{q \times q}} \hspace{0.1cm}\omega\mathrm{(t)}\mathrm{\widetilde{\mathrm{g}}(t)}
\end{equation}
with $\tau>0$ is positive constant and $\tau \mathrm{I_{q \times q}} $ is learning rate matrix. \cite{ortega2020new, roy2019relaxing}. 
The gradient Parametric Error Equation (PEE)
\begin{equation}\label{5}
    \dt{\widetilde{\vartheta}}=-\tau \mathrm{I_{q \times q}} \hspace{0.1cm}\omega\mathrm{(t)}(\mathrm{g}- \omega^{\mathrm{T}}\mathrm{(t)}\hat{\vartheta})=-\tau \mathrm{I_{q \times q}} \hspace{0.1cm}\omega\mathrm{(t)} \omega^{\mathrm{T}}\mathrm{(t)}\widetilde{\vartheta}
\end{equation}
which is a Linear Time Variant (LTV) system. 

\subsection{Persistence of Excitation }
The solution 
\begin{equation}\label{16}
    \mathrm{\widetilde{\vartheta}(t)}=exp\left [ -\tau \mathrm{I_{q \times q}}\int_{t_0}^{t} \mathrm{\omega(\tau)\omega^{T}(\tau) d\tau}\right ]\mathrm{\widetilde{\vartheta}_0}, \hspace{0.3cm} \mathrm{\widetilde{\vartheta}(t_0)=\widetilde{\vartheta}_0}
\end{equation}
of (\ref{5}) can be shown exponentially convergent to zero 
\begin{itemize}
    \item  by adjusting the learning rate $\tau \mathrm{I_{q \times q}}$ and
    \item by optimizing the input $\mathrm{u(t)}$ to the given system such that $\int_{t_0}^{t} \mathrm{\omega(\tau)\omega^{T}(\tau) d\tau}$ is positive definite.
\end{itemize}

Based on the above observations and discussion, the following data richness conditions are drawn \cite{ ortega2020new, nguyen2018model}.
\begin{defi}\label{defi1}
        A known bounded regressor vector $\mathrm{\omega(t)}$ is persistently exciting, if there exist $\mathrm{T>0}$, $\mathrm{t}\geq \mathrm{t_0}$ and $\rho  \in \mathbb{R}^+$ such that
\begin{equation}\label{17}
    \mathrm{\int_{t}^{t+T}\omega(\tau)\omega^{T}(\tau)d\tau\geq \rho   \mathrm{I_{q \times q}}}.
\end{equation}
    \end{defi}
    \begin{defi}\label{defi2}
        A known bounded regressor vector $\mathrm{\omega(t)}$ is interval exciting over $\mathrm{[t, \hspace{0.1cm}t+T]}$, if there exist $\mathrm{T>0}$, $\mathrm{t\geq t_0}$ and $\rho  \in \mathbb{R}^+$ such that
\begin{equation}\label{18}
   \mathrm{ \int_{t}^{t+T}\omega(\tau)\omega^{T}(\tau)d\tau\geq \rho  \mathrm{I_{q \times q}}}.
\end{equation}
    \end{defi}
    \begin{remark}
        Thus the accurate estimation of parameters is directly related to the excitation conditions mentioned in Definition \ref{defi1} and Definition \ref{defi2}, in which the accuracy and uniqueness of the identified parameters are dependent on data informativity and data sufficiency, i.e., data richness \cite{ shadab2022persistence}.
    \end{remark}

The following proposition \ref{prop1} is a benchmark for systems theory and can be found in research articles and books of all identification, parameter estimation, and adaptive control-related research work \cite{ortega2020new}.  It is based on the theory of gradient descent algorithm \cite{ nguyen2018model}. The partial proof of Proposition \ref{prop1} is already explained in the above Subsections.
\begin{prop}\label{prop1}
    The gradient-based update law
    \begin{equation}\label{19}
    \dt{\hat{\vartheta}}=\tau \mathrm{I_{q \times q}} \hspace{0.1cm}\omega\mathrm{(t)}(\mathrm{g}(t)- \omega^{\mathrm{T}}\hat{\vartheta})
\end{equation}
 for an LRE (\ref{1}) guarantees the following:
\begin{itemize}
    \item The norm of the parameter error vector  is monotonically non-increasing. Mathematically, it is written as:
    \begin{equation}\label{20}
        \left | \widetilde{\vartheta}\mathrm{(t_2)} \right |\leq \left | \widetilde{\vartheta}\mathrm{(t_1)} \right |, \hspace{0.7cm} \forall \mathrm{(t_2)}\geq \mathrm{(t_1)} \in \mathbb{R}_{\geq 0}.
    \end{equation}
    \item The zero equilibrium of the PEE 
    \begin{equation}\label{21}
    \dt{\widetilde{\vartheta}}=-\tau \mathrm{I_{q \times q}} \hspace{0.1cm}\omega\mathrm{(t)} \omega^{\mathrm{T}}\mathrm{(t)}\widetilde{\vartheta}
\end{equation}    
    is globally exponentially stable (GES) \textit{if and only if} $\omega\mathrm{(t)}\in \mathrm{PE}$.
    \item There exists an optimal value of learning rate $\tau$ for which the parametric error $\widetilde{\vartheta}$ 
     converges to zero with the highest rate of convergence.
\end{itemize}
\end{prop}
\begin{remark}\label{remark3}
   The parameters are estimated such that the output exactly approximates the real one, but these estimates will have a larger variation in relation to the actual value when the regressor $\omega\mathrm{(t)}\notin$ PE.
\end{remark}
The actual visualization of the above Remark \ref{remark3} is specified in online monitoring of transformer \cite{shadab2023finite}. 
\begin{remark}\label{remark4}
    Selecting the learning rate in GE is a daunting task. A higher learning rate results in a larger step size, which raises the possibility that GE may miss the global minima.
\end{remark}
 Some examples are presented in \cite{aranovskiy2016parameters, ortega2020new, shadab2023finite} to get a clear understanding of the Remark \ref{remark3} and Remark \ref{remark4}. PE is an extremely restrictive and stringent condition, hence relaxing it is preferred for system control, analysis, and performance enhancement. In order to relax the PE condition, different parameter estimators are proposed in the literature.


\section{Problem Formulation and Statement of Research}\label{pf}
The underlying structure of any system dynamics can be made a problem of online estimation of unknown parameter vector appearing linearly with regressor vector as:
\begin{equation}\label{22}
\mathrm{g}=\mathrm{\omega^T(t)}\hspace{0.075cm}\vartheta
\end{equation}
where $\mathrm{g}:\mathbb{R}^{\mathrm{q}}\rightarrow \mathbb{R}$ represents the output form of a given system model, $\mathrm{\omega(t)}\in \mathbb{R}^{\mathrm{q}}$ is known regressor vector, and $\vartheta \in\mathbb{R}^{\mathrm{q}} $ are actual parameters appearing linearly. 
The goal is to construct an estimate $\hat{\vartheta}$ of $\vartheta$, such that the parametric error  $\widetilde{\vartheta}=\vartheta-\hat{\vartheta}$ converges to zero, i.e., 
\begin{equation}\label{23}
    \underset{\mathrm{t}\rightarrow \infty}{lim}\left \| \widetilde{\vartheta}\mathrm{(t)} \right \|=0.
\end{equation}

The issues and limitations of these recent parameter estimation procedures are outlined below:
\begin{itemize}
    \item Choosing the learning rate in GE is tricky and complicated. It must be as low as possible. A larger step size results from a higher learning rate, which increases the chance that GE may miss the global minima \cite{ShadabArxiv}.
    \item In CL, the derivation of all the system states is required for calculating prediction errors which may not be always feasible.
    \item Contrary to CL, a second-order filter is implemented in composite learning to address the system state derivative problem. One of the parameters is the integration window length, which is not always calculable in advance.
    \item The original LRE (\ref{1}) is extended in very first step of the DREM to obtain the extended LREs using linear stable filters. In order to get the $\mathrm{q \times q}$ square matrix, this procedure requires $\mathrm{q}-1$ filters for $\mathrm{q}$-parameters. The major concern is how to carry out such an extension while maintaining the regressor's excitation level. Even if the original regressor $\omega \mathrm{(t)}$ is PE, a poor choice can compromise the convergence. Moreover, the improper tuning of  the coefficient values of filters and learning rate matrix hampers the effective implementation of DREM in certain applications \cite{bobtsov2022generation}.
    \item Multiple examples \cite{bobtsov2022generation} have demonstrated that consistent parameter estimate of a linear time-invariant (LTI) system with DREM is only achievable if the original regressor is PE \cite{bobtsov2022generation}. 
    \item The improper tuning of  the coefficient values of filters may lead to rank-deficient matrix in DREM. The step of DREM that comprises the matrix inversion may fail. As the new regressor is based on the  determinant of an extended regression matrix, this may produce the erroneous calculations and improper new extended regressor.
\end{itemize}

\section{Proposed Passivity and Immersion-based Modified Gradient Estimator}
Concerning the issues and challenges associated with the aforementioned recent parameter estimation procedures, a constructive and systematic strategy with more apparent degrees of freedom to achieve the accurate estimation of unknown parameters via a control perspective is proposed. By adding a virtual control in the final equation of the gradient dynamics, the proposed passivity and immersion-based modified gradient estimator (MGE) enhance and modify the general gradient estimator. The solution of the virtual control law is identified by the P\&I approach \cite{ShadabArxiv}. The P\&I approach \cite{ShadabArxiv} is based on the choice of an appropriate implicit manifold and the generation of a suitable passive output and a related storage function. This facilitates the virtual control law being obtained in a way that the parametric error converges exponentially to zero. Therefore, this paper attempts to sketch the outline of a proposed theory based on the control perspective which provides a stepping stone to be general enough. The proposed P\&I-based modified gradient estimator is extended via the MRE approach.

The equation (\ref{1}) is re-written as:
\begin{equation}\label{24}
\mathrm{g}=\mathrm{\omega_1(t)}\vartheta_1+\mathrm{\omega_2(t)}\vartheta_2+\mathrm{\omega_3(t)}\vartheta_3+......+\mathrm{\omega_{q-1}(t)}\vartheta_{\mathrm{q-1}}+\mathrm{\omega_q(t)}\vartheta_{\mathrm{q}}
\end{equation}
The GE (\ref{4}) for (\ref{24}) is given by:
\begin{align}\label{25}
\begin{split}
& \dt{\hat{\vartheta}}_1=\tau \hspace{0.1cm}\omega_1\mathrm{(t)}\left ( \mathrm{g}(t)-\left ( \mathrm{\omega_1(t)}\hat{\vartheta}_1+\mathrm{\omega_2(t)}\hat{\vartheta}_2+\mathrm{\omega_3(t)}\hat{\vartheta}_3+......+\mathrm{\omega_{q-1}(t)}\hat{\vartheta}_{\mathrm{q-1}}+\mathrm{\omega_q(t)}\hat{\vartheta}_{\mathrm{q}} \right ) \right )\\
&\dt{\hat{\vartheta}}_2=\tau \hspace{0.1cm}\omega_2\mathrm{(t)}\left ( \mathrm{g}(t)-\left ( \mathrm{\omega_1(t)}\hat{\vartheta}_1+\mathrm{\omega_2(t)}\hat{\vartheta}_2+\mathrm{\omega_3(t)}\hat{\vartheta}_3+......+\mathrm{\omega_{q-1}(t)}\hat{\vartheta}_{\mathrm{q-1}}+\mathrm{\omega_q(t)}\hat{\vartheta}_{\mathrm{q}} \right ) \right )\\
&\vdots \hspace{1cm}\vdots \hspace{1cm}\vdots \hspace{1cm}\vdots \hspace{1cm}\vdots \hspace{1cm}\vdots \hspace{1cm}\vdots \hspace{1cm}\vdots \hspace{1cm}\vdots \hspace{1cm}\vdots \hspace{1cm}\vdots \hspace{1cm}\vdots  \\
&\dt{\hat{\vartheta}}_{\mathrm{q-1}}=\tau \hspace{0.1cm}\omega_{\mathrm{q-1}}\mathrm{(t)}\left ( \mathrm{g}(t)-\left ( \mathrm{\omega_1(t)}\hat{\vartheta}_1+\mathrm{\omega_2(t)}\hat{\vartheta}_2+\mathrm{\omega_3(t)}\hat{\vartheta}_3+......+\mathrm{\omega_{q-1}(t)}\hat{\vartheta}_{\mathrm{q-1}}+\mathrm{\omega_q(t)}\hat{\vartheta}_{\mathrm{q}} \right ) \right )\\
&\dt{\hat{\vartheta}}_{\mathrm{q}}=\tau \hspace{0.1cm}\omega_{\mathrm{q}}\mathrm{(t)}\left ( \mathrm{g}(t)-\left ( \mathrm{\omega_1(t)}\hat{\vartheta}_1+\mathrm{\omega_2(t)}\hat{\vartheta}_2+\mathrm{\omega_3(t)}\hat{\vartheta}_3+......+\mathrm{\omega_{q-1}(t)}\hat{\vartheta}_{\mathrm{q-1}}+\mathrm{\omega_q(t)}\hat{\vartheta}_{\mathrm{q}} \right ) \right )
    \end{split}
\end{align}
The parametric error equation (PEE) (\ref{5}) 
\begin{align}\label{26}
\begin{split}
& \dt{\widetilde{\vartheta}}_1=-\tau \hspace{0.1cm}\omega_1\mathrm{(t)}\left ( \mathrm{g}(t)-\left ( \mathrm{\omega_1(t)}\widetilde{\vartheta}_1+\mathrm{\omega_2(t)}\widetilde{\vartheta}_2+\mathrm{\omega_3(t)}\widetilde{\vartheta}_3+......+\mathrm{\omega_{q-1}(t)}\widetilde{\vartheta}_{\mathrm{q-1}}+\mathrm{\omega_q(t)}\widetilde{\vartheta}_{\mathrm{q}} \right ) \right )\\
&\dt{\widetilde{\vartheta}}_2=-\tau \hspace{0.1cm}\omega_2\mathrm{(t)}\left ( \mathrm{g}(t)-\left ( \mathrm{\omega_1(t)}\widetilde{\vartheta}_1+\mathrm{\omega_2(t)}\widetilde{\vartheta}_2+\mathrm{\omega_3(t)}\widetilde{\vartheta}_3+......+\mathrm{\omega_{q-1}(t)}\widetilde{\vartheta}_{\mathrm{q-1}}+\mathrm{\omega_q(t)}\widetilde{\vartheta}_{\mathrm{q}} \right ) \right )\\
&\vdots \hspace{1cm}\vdots \hspace{1cm}\vdots \hspace{1cm}\vdots \hspace{1cm}\vdots \hspace{1cm}\vdots \hspace{1cm}\vdots \hspace{1cm}\vdots \hspace{1cm}\vdots \hspace{1cm}\vdots \hspace{1cm}\vdots \hspace{1cm}\vdots  \\
&\dt{\widetilde{\vartheta}}_{\mathrm{q-1}}=-\tau \hspace{0.1cm}\omega_{\mathrm{q-1}}\mathrm{(t)}\left ( \mathrm{g}(t)-\left ( \mathrm{\omega_1(t)}\widetilde{\vartheta}_1+\mathrm{\omega_2(t)}\widetilde{\vartheta}_2+\mathrm{\omega_3(t)}\widetilde{\vartheta}_3+......+\mathrm{\omega_{q-1}(t)}\widetilde{\vartheta}_{\mathrm{q-1}}+\mathrm{\omega_q(t)}\widetilde{\vartheta}_{\mathrm{q}} \right ) \right )\\
&\dt{\widetilde{\vartheta}}_{\mathrm{q}}=-\tau \hspace{0.1cm}\omega_{\mathrm{q}}\mathrm{(t)}\left ( \mathrm{g}(t)-\left ( \mathrm{\omega_1(t)}\widetilde{\vartheta}_1+\mathrm{\omega_2(t)}\widetilde{\vartheta}_2+\mathrm{\omega_3(t)}\widetilde{\vartheta}_3+......+\mathrm{\omega_{q-1}(t)}\widetilde{\vartheta}_{\mathrm{q-1}}+\mathrm{\omega_q(t)}\widetilde{\vartheta}_{\mathrm{q}} \right ) \right )
    \end{split}
\end{align}
is explicitly modified as 
\begin{align}\label{27}
    \begin{split}
 &\dt{\widetilde{\vartheta}}_1=-\tau\hspace{0.1cm}\omega_1\mathrm{(t)}\omega_1\mathrm{(t)}  \widetilde{\vartheta}_1-\tau\omega_1\mathrm{(t)}\omega_2\mathrm{(t)}  \widetilde{\vartheta}_2-.....-\tau\omega_1\mathrm{(t)}\omega_{\mathrm{q-1}}\mathrm{(t)}  \widetilde{\vartheta}_{\mathrm{q-1}}-\tau\omega_1\mathrm{(t)}\omega_{\mathrm{q}}\mathrm{(t)}  \widetilde{\vartheta}_{\mathrm{q}}\\
&\dt{\widetilde{\vartheta}}_2=-\tau\hspace{0.1cm}\omega_2\mathrm{(t)}\omega_1\mathrm{(t)}  \widetilde{\vartheta}_1-\tau\omega_2\mathrm{(t)}\omega_2\mathrm{(t)}  \widetilde{\vartheta}_2-.....-\tau\omega_2\mathrm{(t)}\omega_{\mathrm{q-1}}\mathrm{(t)}  \widetilde{\vartheta}_{\mathrm{q-1}}-\tau\omega_2\mathrm{(t)}\omega_{\mathrm{q}}\mathrm{(t)}  \widetilde{\vartheta}_{\mathrm{q}}\\
&\vdots \hspace{1cm} \vdots \hspace{1cm} \vdots \hspace{1cm} \vdots \hspace{1cm} \vdots \hspace{1cm} \vdots \hspace{1cm} \vdots \hspace{1cm} \vdots \hspace{1cm} \vdots \hspace{1cm} \vdots \\
&\dt{\widetilde{\vartheta}}_{\mathrm{q-1}}=-\tau\hspace{0.1cm}\mathrm{(t)}\omega_{\mathrm{q-1}}\mathrm{(t)}\omega_1\mathrm{(t)}  \widetilde{\vartheta}_1-\tau\omega_{\mathrm{q-1}}\mathrm{(t)}\omega_2\mathrm{(t)}  \widetilde{\vartheta}_2-.....-\tau\omega_{\mathrm{q-1}}\mathrm{(t)}\omega_{\mathrm{q-1}}\mathrm{(t)}  \widetilde{\vartheta}_{\mathrm{q-1}}-\tau\omega_{\mathrm{q-1}}\mathrm{(t)}\omega_{\mathrm{q}}\mathrm{(t)}  \widetilde{\vartheta}_{\mathrm{q}}\\
&\dt{\widetilde{\vartheta}}_{\mathrm{q}}=-\tau\hspace{0.1cm}\omega_{\mathrm{q}}\mathrm{(t)}\omega_{\mathrm{q}}\mathrm{(t)}\omega_1\mathrm{(t)}  \widetilde{\vartheta}_1-\tau\omega_{\mathrm{q}}\omega_2\mathrm{(t)}  \widetilde{\vartheta}_2-.....-\tau\omega_{\mathrm{q}}\mathrm{(t)}\omega_{\mathrm{q-1}}\mathrm{(t)}  \widetilde{\vartheta}_{\mathrm{q-1}}-\tau\omega_{\mathrm{q}}\mathrm{(t)}\omega_{\mathrm{q}}\mathrm{(t)}  \widetilde{\vartheta}_{\mathrm{q}}+\mathbb{\delta_{\mathrm{u}}}
    \end{split}
\end{align}
by adding a virtual control law $\mathbb{\delta_{\mathrm{u}}}$ in the last equation i.e., added to $\dt{\widetilde{\vartheta}}_{\mathrm{q}}$. 
\begin{remark}
    The addition of $\mathbb{\delta_{\mathrm{u}}}$ in the $\dt{\widetilde{\vartheta}}_{\mathrm{q}}$ is the major contribution this paper. The idea of stabilization and control is utilized in this P\&I-based gradient estimator by externally adding the virtual control term $\mathbb{\delta_{\mathrm{u}}}$. The proposed approach is implemented in a closed-loop form, in contrast to well-known open-loop parameter estimators.
\end{remark}
\begin{remark}
    The goal is to obtain the value of $\mathbb{\delta_{\mathrm{u}}}$ such that the solution of the complete PEE-dynamics (\ref{27}) converges exponentially/asymptotically to zero. As the $\widetilde{\vartheta}=\vartheta-\hat{\vartheta}$ is considered, the convergence of $\widetilde{\vartheta}\rightarrow 0$ implies $\hat{\vartheta}\rightarrow \vartheta$.
\end{remark}
The parameter estimation problem can be visualized as a control and stabilization problem. Therefore, the recently developed P\&I approach \cite{ShadabArxiv} for control and stabilization (Detail explanation and procedure are provided in Appendix \ref{Appendix 2}) is applied to get the $\mathbb{\delta_{\mathrm{u}}}$.
The P\&I approach basically involves four steps procedure. The very first step is the construction of the implicit manifold.
Therefore, the target dynamics
\begin{equation}\label{28}
\dot{\eta}=\beta(\eta)=-\tau\hspace{0.1cm}\omega_1\mathrm{(t)}\omega_1\mathrm{(t)} \eta-\tau\omega_1\mathrm{(t)}\omega_2\mathrm{(t)} \mu \eta-.....-\tau\omega_1\mathrm{(t)}\omega_{\mathrm{q-1}}\mathrm{(t)} \mu  \eta-\tau\omega_1\mathrm{(t)}\omega_{\mathrm{q}}\mathrm{(t)} \mu  \eta
\end{equation}
with $\widetilde{\vartheta}_1=\eta$ is defined. This defines the implicit manifolds  $\Psi_1(\widetilde{\vartheta}_1, \widetilde{\vartheta}_2)=\widetilde{\vartheta}_2-\mu \widetilde{\vartheta}_1=0$, $\Psi_2(\widetilde{\vartheta}_1, \widetilde{\vartheta}_3)=\widetilde{\vartheta}_3-\mu \widetilde{\vartheta}_1=0$,......, $\Psi_{\mathrm{q-1}}(\widetilde{\vartheta}_1, \widetilde{\vartheta}_{\mathrm{q-1}})=\widetilde{\vartheta}_{\mathrm{q-1}}-\mu \widetilde{\vartheta}_1=0$, and $\Psi_{\mathrm{q}}(\widetilde{\vartheta}_1, \widetilde{\vartheta}_{\mathrm{q-1}})=\widetilde{\vartheta}_{\mathrm{q}}-\mu \widetilde{\vartheta}_1=0$.
The addition of all these implicit manifolds provides a combined implicit manifold given as:
\begin{equation}\label{29}
    \Phi(\widetilde{\vartheta})=\left ( \left ( \sum_{\mathrm{i=1}}^{\mathrm{q-1}}\widetilde{\vartheta}_{\mathrm{i+1}} \right )-(\mathrm{q-1})\mu \widetilde{\vartheta}_1\right)=\widetilde{\vartheta}_{\mathrm{q}}+\widetilde{\vartheta}_{\mathrm{q-1}}+\widetilde{\vartheta}_{\mathrm{q-3}}+.....+\widetilde{\vartheta}_{\mathrm{2}}-(\mathrm{q-1})\mu 
 \widetilde{\vartheta}_{\mathrm{1}}
\end{equation}
The application of $S_2$, $S_3$, and $S_4$ of the P\&I approach (Detail explanation in Appendix \ref{Appendix 2}) provides the control law 
\begin{align}\label{30}
    \begin{split}  \mathbb{\delta_{\mathrm{u}}}=&\tau\hspace{0.1cm}\omega_{\mathrm{q}}\mathrm{(t)}\omega_{\mathrm{q}}\mathrm{(t)}\omega_1\mathrm{(t)}  \widetilde{\vartheta}_1+\tau\omega_{\mathrm{q}}\omega_2\mathrm{(t)}  \widetilde{\vartheta}_2+.....+\tau\omega_{\mathrm{q}}\mathrm{(t)}\omega_{\mathrm{q-1}}\mathrm{(t)}  \widetilde{\vartheta}_{\mathrm{q-1}}+\tau\omega_{\mathrm{q}}\mathrm{(t)}\omega_{\mathrm{q}}\mathrm{(t)}  \widetilde{\vartheta}_{\mathrm{q}}
    \tau\hspace{0.1cm}\mathrm{(t)}\omega_{\mathrm{q-1}}\mathrm{(t)}\omega_1\mathrm{(t)}  \widetilde{\vartheta}_1\\&+\tau\omega_{\mathrm{q-1}}\mathrm{(t)}\omega_2\mathrm{(t)}  \widetilde{\vartheta}_2+.....+\tau\omega_{\mathrm{q-1}}\mathrm{(t)}\omega_{\mathrm{q-1}}\mathrm{(t)}  \widetilde{\vartheta}_{\mathrm{q-1}}+\tau\omega_{\mathrm{q-1}}\mathrm{(t)}\omega_{\mathrm{q}}\mathrm{(t)}  \widetilde{\vartheta}_{\mathrm{q}}
+.........+\tau\hspace{0.1cm}\omega_2\mathrm{(t)}\omega_1\mathrm{(t)}  \widetilde{\vartheta}_1+\\&\tau\omega_2\mathrm{(t)}\omega_2\mathrm{(t)}  \widetilde{\vartheta}_2+.....+\tau\omega_2\mathrm{(t)}\omega_{\mathrm{q-1}}\mathrm{(t)}  \widetilde{\vartheta}_{\mathrm{q-1}}+\tau\omega_2\mathrm{(t)}\omega_{\mathrm{q}}\mathrm{(t)}  \widetilde{\vartheta}_{\mathrm{q}}-\tau\mu (\mathrm{q-1})\hspace{0.1cm}\omega_1\mathrm{(t)}\omega_1\mathrm{(t)}  \widetilde{\vartheta}_1\\&-\tau\mu (\mathrm{q-1})\omega_1\mathrm{(t)}\omega_2\mathrm{(t)}  \widetilde{\vartheta}_2-.....-\tau\mu (\mathrm{q-1})\omega_1\mathrm{(t)}\omega_{\mathrm{q-1}}\mathrm{(t)}  \widetilde{\vartheta}_{\mathrm{q-1}}-\tau\mu (\mathrm{q-1})\omega_1\mathrm{(t)}\omega_{\mathrm{q}}\mathrm{(t)}  \widetilde{\vartheta}_{\mathrm{q}}\\&-\frac{\alpha}{2}(\widetilde{\vartheta}_{\mathrm{q}}+\widetilde{\vartheta}_{\mathrm{q-1}}+\widetilde{\vartheta}_{\mathrm{q-3}}+.....+\widetilde{\vartheta}_{\mathrm{2}}-(\mathrm{q-1})\mu 
 \widetilde{\vartheta}_{\mathrm{1}}).
    \end{split}
\end{align}
\begin{remark}
    As the calculation of the term $-\frac{\alpha}{2}(\widetilde{\vartheta}_{\mathrm{q}}+\widetilde{\vartheta}_{\mathrm{q-1}}+\widetilde{\vartheta}_{\mathrm{q-3}}+.....+\widetilde{\vartheta}_{\mathrm{2}}-(\mathrm{q-1})\mu 
 \widetilde{\vartheta}_{\mathrm{1}})$ required the actual value, the value $\alpha=0$ is chosen. This leads to the $\dt{\mathbb{S}}\leq 0$.
\end{remark}
The substitution of equation (\ref{30})  in (\ref{27}) ensures the asymptotic stability of zero equilibrium  of CLS. 
The final form of the proposed MGE is given by:
\begin{align}\label{125}
\begin{split}
& \dt{\hat{\vartheta}}_1=\tau \hspace{0.1cm}\omega_1\mathrm{(t)}\left ( \mathrm{g}(t)-\left ( \mathrm{\omega_1(t)}\hat{\vartheta}_1+\mathrm{\omega_2(t)}\hat{\vartheta}_2+\mathrm{\omega_3(t)}\hat{\vartheta}_3+......+\mathrm{\omega_{q-1}(t)}\hat{\vartheta}_{\mathrm{q-1}}+\mathrm{\omega_q(t)}\hat{\vartheta}_{\mathrm{q}} \right ) \right )\\
&\dt{\hat{\vartheta}}_2=\tau \hspace{0.1cm}\omega_2\mathrm{(t)}\left ( \mathrm{g}(t)-\left ( \mathrm{\omega_1(t)}\hat{\vartheta}_1+\mathrm{\omega_2(t)}\hat{\vartheta}_2+\mathrm{\omega_3(t)}\hat{\vartheta}_3+......+\mathrm{\omega_{q-1}(t)}\hat{\vartheta}_{\mathrm{q-1}}+\mathrm{\omega_q(t)}\hat{\vartheta}_{\mathrm{q}} \right ) \right )\\
&\vdots \hspace{1cm}\vdots \hspace{1cm}\vdots \hspace{1cm}\vdots \hspace{1cm}\vdots \hspace{1cm}\vdots \hspace{1cm}\vdots \hspace{1cm}\vdots \hspace{1cm}\vdots \hspace{1cm}\vdots \hspace{1cm}\vdots \hspace{1cm}\vdots  \\
&\dt{\hat{\vartheta}}_{\mathrm{q-1}}=\tau \hspace{0.1cm}\omega_{\mathrm{q-1}}\mathrm{(t)}\left ( \mathrm{g}(t)-\left ( \mathrm{\omega_1(t)}\hat{\vartheta}_1+\mathrm{\omega_2(t)}\hat{\vartheta}_2+\mathrm{\omega_3(t)}\hat{\vartheta}_3+......+\mathrm{\omega_{q-1}(t)}\hat{\vartheta}_{\mathrm{q-1}}+\mathrm{\omega_q(t)}\hat{\vartheta}_{\mathrm{q}} \right ) \right )\\
&\dot{\hat{\vartheta}}_{\mathrm{q}}=\delta_{\mathrm{u}}-\dot{\widetilde{\vartheta}}_{\mathrm{q}}=\delta_{\mathrm{u}}+(\tau \hspace{0.1cm}\omega_{\mathrm{q}}\mathrm{(t)}\left ( \mathrm{g}(t)-\left ( \mathrm{\omega_1(t)}\hat{\vartheta}_1+\mathrm{\omega_2(t)}\hat{\vartheta}_2+\mathrm{\omega_3(t)}\hat{\vartheta}_3+......+\mathrm{\omega_{q-1}(t)}\hat{\vartheta}_{\mathrm{q-1}}+\mathrm{\omega_q(t)}\hat{\vartheta}_{\mathrm{q}} \right ) \right ))
    \end{split}
\end{align}
\textbf{The proposed P\&I based MGE for $\mathrm{q}=2$:}
The equation (\ref{1}) for $\mathrm{q}=2$ is written as:
\begin{equation}\label{231}
\mathrm{g}=\mathrm{\omega_1(t)}\vartheta_1+\mathrm{\omega_2(t)}\vartheta_2
\end{equation}
\begin{align}\label{232}
\begin{split}
& \dt{\hat{\vartheta}}_1=\tau \hspace{0.1cm}\omega_1\mathrm{(t)}\left ( \mathrm{g}(t)-\left ( \mathrm{\omega_1(t)}\hat{\vartheta}_1+\mathrm{\omega_2(t)}\hat{\vartheta}_2 \right ) \right )\\
&\dt{\hat{\vartheta}}_2=\tau \hspace{0.1cm}\omega_2\mathrm{(t)}\left ( \mathrm{g}(t)-\left ( \mathrm{\omega_1(t)}\hat{\vartheta}_1+\mathrm{\omega_2(t)}\hat{\vartheta}_2 \right ) \right ).
    \end{split}
\end{align}
The parametric error equation (PEE) (\ref{5}) 
\begin{align}\label{233}
\begin{split}
& \dt{\widetilde{\vartheta}}_1=-\tau \hspace{0.1cm}\omega_1\mathrm{(t)}\left ( \mathrm{g}(t)-\left ( \mathrm{\omega_1(t)}\hat{\vartheta}_1+\mathrm{\omega_2(t)}\hat{\vartheta}_2 \right ) \right )\\
&\dt{\widetilde{\vartheta}}_2=-\tau \hspace{0.1cm}\omega_2\mathrm{(t)}\left ( \mathrm{g}(t)-\left ( \mathrm{\omega_1(t)}\hat{\vartheta}_1+\mathrm{\omega_2(t)}\hat{\vartheta}_2\right ) \right )
    \end{split}
\end{align}
is explicitly modified as 
\begin{align}\label{234}
\begin{split}
& \dt{\widetilde{\vartheta}}_1=-\tau \hspace{0.1cm}\omega_1\mathrm{(t)} \left ( \mathrm{\omega_1(t)}\widetilde{\vartheta}_1+\mathrm{\omega_2(t)}\widetilde{\vartheta}_2 \right ) \\
&\dt{\widetilde{\vartheta}}_2=-\tau \hspace{0.1cm}\omega_2\mathrm{(t)} \left ( \mathrm{\omega_1(t)}\widetilde{\vartheta}_1+\mathrm{\omega_2(t)}\widetilde{\vartheta}_2\right ) +\delta_{\mathrm{u}}
    \end{split}
\end{align}
by adding a virtual control law $\mathbb{\delta_{\mathrm{u}}}$ in the last equation i.e., added to $\dt{\widetilde{\vartheta}}_{\mathrm{2}}$. The recently developed P\&I approach \cite{ShadabArxiv} for control and stabilization is applied to get the $\mathbb{\delta_{\mathrm{u}}}$. The very first step is the construction of the implicit manifold.
Therefore, the target dynamics $\dt{\eta}=-\tau\hspace{0.1cm}\omega_1\mathrm{(t)}\omega_1\mathrm{(t)}  \eta -\tau\mu \omega_1\mathrm{(t)}\omega_2\mathrm{(t)}  \eta$ with $\widetilde{\vartheta}_1=\eta$ is selected. This helps in defining the implicit manifold 
\begin{equation}\label{235}
    \widetilde{\vartheta}_2-\mu \widetilde{\vartheta}_1=0.
\end{equation}
The application of four step procedure $(S_1-S_4)$ of the P\&I approach with $\alpha=0$ provides the virtual control law
\begin{equation}\label{236}
    \delta_{\mathrm{u}}=(\tau\omega_2-\mu \tau\omega_1)\left ( \mathrm{\omega_1(t)}\widetilde{\vartheta}_1+\mathrm{\omega_2(t)}\widetilde{\vartheta}_2 \right ) =(\tau\omega_2-\mu \tau\omega_1)\left ( \mathrm{g}(t)-\left ( \mathrm{\omega_1(t)}\hat{\vartheta}_1+\mathrm{\omega_2(t)}\hat{\vartheta}_2 \right ) \right )
\end{equation}
The last equation 
$\dt{\widetilde{\vartheta}}_{\mathrm{2}}=-\tau\hspace{0.1cm}\omega_{\mathrm{2}}\mathrm{(t)}\omega_1\mathrm{(t)}  \widetilde{\vartheta}_1-\tau\omega_{\mathrm{2}}\mathrm{(t)}\omega_2\mathrm{(t)}  \widetilde{\vartheta}_2+\delta_{\mathrm{u}}$ of (\ref{234}) is re-written as $\dt{\widetilde{\vartheta}}_{\mathrm{2}}=-\tau\hspace{0.1cm}\omega_{\mathrm{2}}\left (\mathrm{g}-\mathrm{\omega_1(t)}\hat{\vartheta}_1-\mathrm{\omega_2(t)}\hat{\vartheta}_2 \right )+\delta_{\mathrm{u}}=-\dt{\hat{\vartheta}}_2+\delta_{\mathrm{u}}$. Based on this mathematical calculation, the estimation dynamics $\dt{\hat{\vartheta}}_3$ is written as:
\begin{equation}\label{237}
  \dt{\hat{\vartheta}}_2=\delta_{\mathrm{u}}-\dt{\widetilde{\vartheta}}_{\mathrm{2}}=(2\tau\omega_2-\mu\tau\omega_1)\left (\mathrm{g}-\mathrm{\omega_1(t)}\hat{\vartheta}_1-\mathrm{\omega_2(t)}\hat{\vartheta}_2 \right ).
\end{equation}
The corresponding error equation is explicitly written as:
\begin{equation}\label{238}
    \dt{\widetilde{\vartheta}}_2=-(2\tau\omega_2-\mu\tau\omega_1)\left ( \hspace{0.1cm}\omega_1\mathrm{(t)}  \widetilde{\vartheta}_1+\omega_2\mathrm{(t)}  \widetilde{\vartheta}_2 \right ).
\end{equation}

With this (\ref{237}), The MGE is written as:
\begin{align}\label{239}
\begin{split}
&  \dt{\hat{\vartheta}}_1=\tau\hspace{0.1cm}\omega_1\mathrm{(t)}\left (\mathrm{g}-\mathrm{\omega_1(t)}\hat{\vartheta}_1-\mathrm{\omega_2(t)}\hat{\vartheta}_2  \right )\\
&\dt{\hat{\vartheta}}_2=\delta_{\mathrm{u}}-\dt{\widetilde{\vartheta}}_{\mathrm{2}}=(2\tau\omega_2-\mu\tau\omega_1)\left (\mathrm{g}-\mathrm{\omega_1(t)}\hat{\vartheta}_1-\mathrm{\omega_2(t)}\hat{\vartheta}_2 \right )
    \end{split}
\end{align}

When the parametric error  $\widetilde{\vartheta}=\vartheta-\hat{\vartheta}$ converges to zero, the accurate parameter estimates $\hat{\vartheta}$ of $\vartheta$ using the  dynamics $\dt{\widetilde{\vartheta}}=-\dt{\hat{\vartheta}}$ is easily obtained. Thus, the final parametric error dynamics of MGE is re-written as:
\begin{align}\label{240}
    \begin{split}
    &\dt{\widetilde{\vartheta}}_2=-\tau\omega_1\left ( \hspace{0.1cm}\omega_1\mathrm{(t)}  \widetilde{\vartheta}_1+\omega_2\mathrm{(t)}  \widetilde{\vartheta}_2 \right )\\
 &\dt{\widetilde{\vartheta}}_2=-(2\tau\omega_2-\mu\tau\omega_1)\left ( \hspace{0.1cm}\omega_1\mathrm{(t)}  \widetilde{\vartheta}_1+\omega_2\mathrm{(t)}  \widetilde{\vartheta}_2 \right ).
    \end{split}
\end{align}

\begin{figure}
    \centering
    \includegraphics[width=\linewidth]{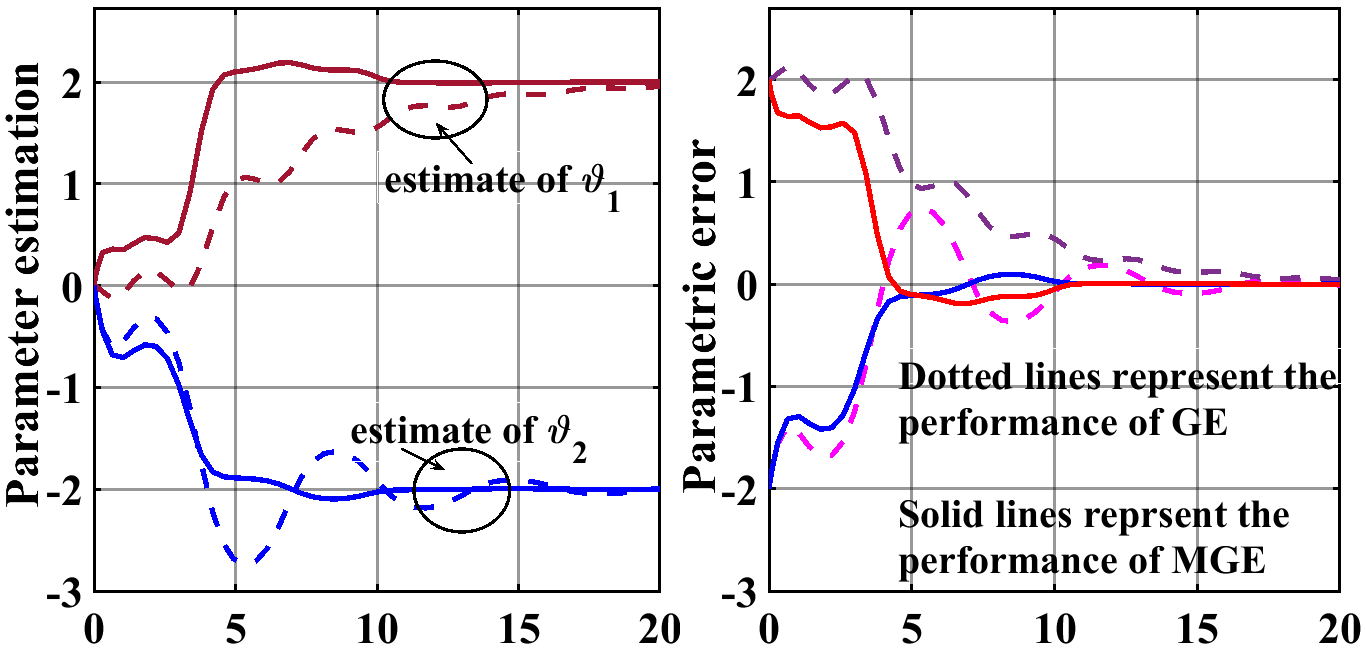}
    \caption{Time Evolution of parameter estimates and parametric error for $\mu=0.95$ and $\tau=1$.}
    \label{fig1}
\end{figure}

\begin{example}\label{example1}
   To illustrate the proposed P\&I based MGE (\ref{239}),  a PE regressor $\omega(\mathrm{t})=\mathrm{col}(1, \mathrm{sin(t)})$ \cite{shadab2023finite} with $\vartheta=\mathrm{col(-2, 2)}$ is considered. Based on the above error dynamics (\ref{240}), the time evolution of parametric error $\widetilde{\vartheta}(\mathrm{t})$ with the transient performance  of the proposed P\&I-based MGE is shown in Fig. \ref{fig1}. 
\end{example}

\begin{figure}
    \centering
    \includegraphics[width=\linewidth]{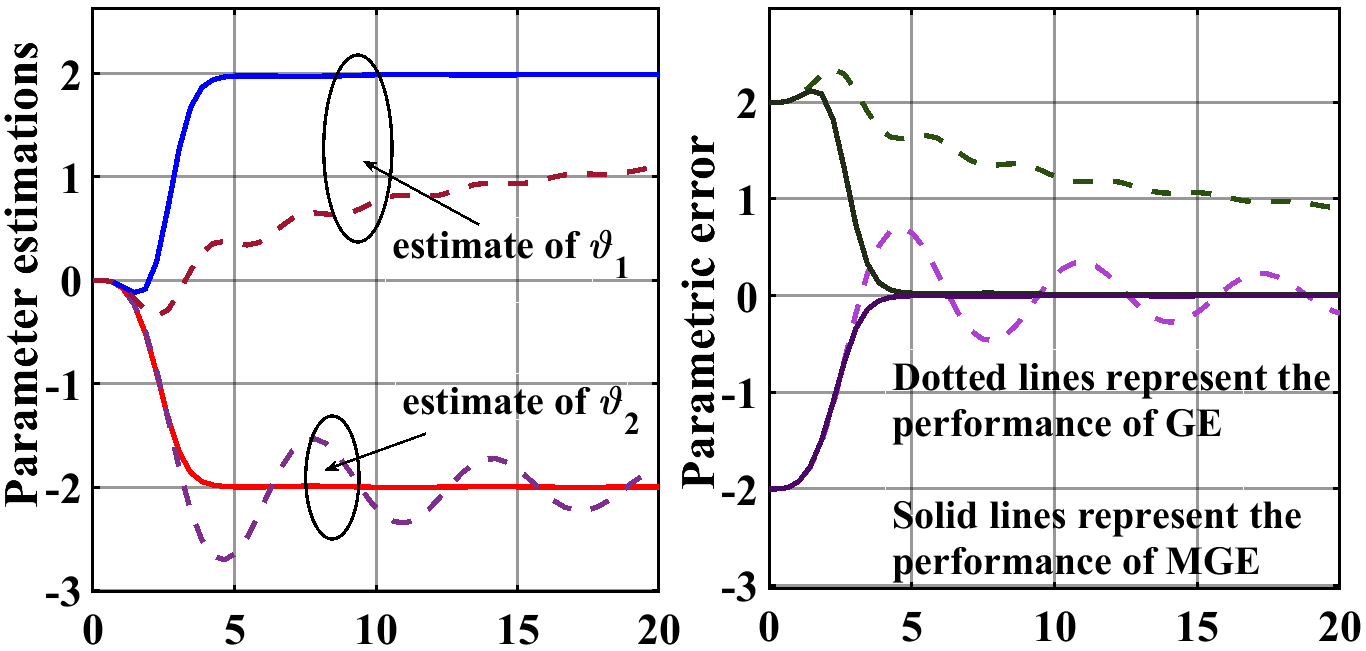}
    \caption{Time Evolution of parameter estimates and parametric error for $\mu=0.95$ and $\tau=1$.}
    \label{fig2}
\end{figure}

\begin{example}\label{example2}
    Consider a not PE regressor 
    \begin{equation}
        \omega(\mathrm{t})=\mathrm{col}(1, \mathrm{((sin(t)+cos(t))/((1+t)^{0.5}))-(sin(t)/((2(1+t)^{1.5})))})
    \end{equation}
     \cite{ortega2020new} with $\vartheta=\mathrm{col(-2, 2)}$. Based on the above error dynamics (\ref{240}), the time evolution of parametric error $\widetilde{\vartheta}(\mathrm{t})$ with the transient performance  of the proposed P\&I-based MGE  (\ref{239}) is shown in Fig. \ref{fig2}. 
\end{example}

\textbf{The proposed P\&I based MGE for $\mathrm{q}=3$:}
The equation (\ref{1}) for $\mathrm{q}=3$ is written as:
\begin{equation}\label{31}
\mathrm{g}=\mathrm{\omega_1(t)}\vartheta_1+\mathrm{\omega_2(t)}\vartheta_2+\mathrm{\omega_3(t)}\vartheta_3
\end{equation}
The GE (\ref{4}) for (\ref{32}) is given by:
\begin{align}\label{32}
\begin{split}
& \dt{\hat{\vartheta}}_1=\tau \hspace{0.1cm}\omega_1\mathrm{(t)}\left ( \mathrm{g}(t)-\left ( \mathrm{\omega_1(t)}\hat{\vartheta}_1+\mathrm{\omega_2(t)}\hat{\vartheta}_2+\mathrm{\omega_3(t)}\hat{\vartheta}_3 \right ) \right )\\
&\dt{\hat{\vartheta}}_2=\tau \hspace{0.1cm}\omega_2\mathrm{(t)}\left ( \mathrm{g}(t)-\left ( \mathrm{\omega_1(t)}\hat{\vartheta}_1+\mathrm{\omega_2(t)}\hat{\vartheta}_2+\mathrm{\omega_3(t)}\hat{\vartheta}_3 \right ) \right )\\
&\dt{\hat{\vartheta}}_{\mathrm{3}}=\tau \hspace{0.1cm}\omega_{\mathrm{3}}\mathrm{(t)}\left ( \mathrm{g}(t)-\left ( \mathrm{\omega_1(t)}\hat{\vartheta}_1+\mathrm{\omega_2(t)}\hat{\vartheta}_2+\mathrm{\omega_3(t)}\hat{\vartheta}_3 \right ) \right ).
    \end{split}
\end{align}

The parametric error equation (PEE) (\ref{5}) 
\begin{align}\label{33}
\begin{split}
& \dt{\widetilde{\vartheta}}_1=-\tau \hspace{0.1cm}\omega_1\mathrm{(t)}\left ( \mathrm{g}(t)-\left ( \mathrm{\omega_1(t)}\widetilde{\vartheta}_1+\mathrm{\omega_2(t)}\widetilde{\vartheta}_2+\mathrm{\omega_3(t)}\widetilde{\vartheta}_3 \right ) \right )\\
&\dt{\widetilde{\vartheta}}_2=-\tau \hspace{0.1cm}\omega_2\mathrm{(t)}\left ( \mathrm{g}(t)-\left ( \mathrm{\omega_1(t)}\widetilde{\vartheta}_1+\mathrm{\omega_2(t)}\widetilde{\vartheta}_2+\mathrm{\omega_3(t)}\widetilde{\vartheta}_3\right ) \right )\\
&\dt{\widetilde{\vartheta}}_{\mathrm{3}}=-\tau \hspace{0.1cm}\omega_{\mathrm{3}}\mathrm{(t)}\left ( \mathrm{g}(t)-\left ( \mathrm{\omega_1(t)}\widetilde{\vartheta}_1+\mathrm{\omega_2(t)}\widetilde{\vartheta}_2+\mathrm{\omega_3(t)}\widetilde{\vartheta}_3 \right ) \right )
    \end{split}
\end{align}
is explicitly modified as  
\begin{align}\label{34}
    \begin{split}
 &\dt{\widetilde{\vartheta}}_1=-\tau\hspace{0.1cm}\omega_1\mathrm{(t)}\omega_1\mathrm{(t)}  \widetilde{\vartheta}_1-\tau\omega_1\mathrm{(t)}\omega_2\mathrm{(t)}  \widetilde{\vartheta}_2-\tau\omega_1\mathrm{(t)}\omega_{\mathrm{3}}\mathrm{(t)}  \widetilde{\vartheta}_{\mathrm{3}}\\
&\dt{\widetilde{\vartheta}}_2=-\tau\hspace{0.1cm}\omega_2\mathrm{(t)}\omega_1\mathrm{(t)}  \widetilde{\vartheta}_1-\tau\omega_2\mathrm{(t)}\omega_2\mathrm{(t)}  \widetilde{\vartheta}_2-\tau\omega_2\mathrm{(t)}\omega_{\mathrm{3}}\mathrm{(t)}  \widetilde{\vartheta}_{\mathrm{3}}\\
&\dt{\widetilde{\vartheta}}_{\mathrm{3}}=-\tau\hspace{0.1cm}\omega_{\mathrm{3}}\mathrm{(t)}\omega_1\mathrm{(t)}  \widetilde{\vartheta}_1-\tau\omega_{\mathrm{3}}\mathrm{(t)}\omega_2\mathrm{(t)}  \widetilde{\vartheta}_2-\tau\omega_{\mathrm{3}}\mathrm{(t)}\omega_{\mathrm{3}}\mathrm{(t)}  \widetilde{\vartheta}_{\mathrm{3}} +\delta_{\mathrm{u}}
    \end{split}
\end{align}
by adding a virtual control law $\mathbb{\delta_{\mathrm{u}}}$ in the last equation i.e., added to $\dt{\widetilde{\vartheta}}_{\mathrm{3}}$. The recently developed P\&I approach \cite{ShadabArxiv} for control and stabilization is applied to get the $\mathbb{\delta_{\mathrm{u}}}$. The very first step is the construction of the implicit manifold.
Therefore, the target dynamics $\dt{\eta}=-\tau\hspace{0.1cm}\omega_1\mathrm{(t)}\omega_1\mathrm{(t)}  \eta -\tau\mu \omega_1\mathrm{(t)}\omega_2\mathrm{(t)}  \eta-\tau\mu \omega_1\mathrm{(t)}\omega_{\mathrm{3}}\mathrm{(t)}  \eta$ with $\widetilde{\vartheta}_1=\eta$ is selected. This helps in defining the implicit manifold 
\begin{equation}\label{35}
    \widetilde{\vartheta}_3+\widetilde{\vartheta}_2-2\mu \widetilde{\vartheta}_1=0.
\end{equation}
The application of four step procedure $(S_1-S_4)$ of the P\&I approach with $\alpha=0$ provides the virtual control law 
\begin{align}\label{36}
    \begin{split}
\delta_{\mathrm{u}}=\tau\omega_3\omega_1\widetilde{\vartheta}_1+\tau\omega_3\omega_2\widetilde{\vartheta}_2&+\tau\omega_3\omega_3\widetilde{\vartheta}_3+\tau\omega_2\omega_1\widetilde{\vartheta}_1+\tau\omega_2\omega_2\widetilde{\vartheta}_2+\tau\omega_2\omega_3\widetilde{\vartheta}_3\\&-2\mu \tau\omega_1\omega_1\widetilde{\vartheta}_1-2\mu \tau\omega_1\omega_2\widetilde{\vartheta}_2-2\mu \tau\omega_1\omega_3\widetilde{\vartheta}_3.
    \end{split}
\end{align}
The last equation 
$\dt{\widetilde{\vartheta}}_{\mathrm{3}}=-\tau\hspace{0.1cm}\omega_{\mathrm{3}}\mathrm{(t)}\omega_1\mathrm{(t)}  \widetilde{\vartheta}_1-\tau\omega_{\mathrm{3}}\mathrm{(t)}\omega_2\mathrm{(t)}  \widetilde{\vartheta}_2-\tau\omega_{\mathrm{3}}\mathrm{(t)}\omega_{\mathrm{3}}\mathrm{(t)}  \widetilde{\vartheta}_{\mathrm{3}} +\delta_{\mathrm{u}}$ of (\ref{36}) is re-written as $\dt{\widetilde{\vartheta}}_{\mathrm{3}}=-\tau\hspace{0.1cm}\omega_{\mathrm{3}}\left (\mathrm{g}-\mathrm{\omega_1(t)}\hat{\vartheta}_1-\mathrm{\omega_2(t)}\hat{\vartheta}_2-\mathrm{\omega_3(t)}\hat{\vartheta}_3  \right )+\delta_{\mathrm{u}}=-\dt{\hat{\vartheta}}_3+\delta_{\mathrm{u}}$. Based on this mathematical calculation, the estimation dynamics $\dt{\hat{\vartheta}}_3$ is written as:
\begin{equation}\label{37}
  \dt{\hat{\vartheta}}_3=\delta_{\mathrm{u}}-\dt{\widetilde{\vartheta}}_{\mathrm{3}}=(2\tau\omega_3+\tau\omega_2-2\mu\tau\omega_1)\left (\mathrm{g}-\mathrm{\omega_1(t)}\hat{\vartheta}_1-\mathrm{\omega_2(t)}\hat{\vartheta}_2-\mathrm{\omega_3(t)}\hat{\vartheta}_3  \right ).
\end{equation}
The corresponding error equation is explicitly written as:
\begin{equation}\label{38}
    \dt{\widetilde{\vartheta}}_3=-(2\tau\omega_3+\tau\omega_2-2\mu\tau\omega_1)\left ( \hspace{0.1cm}\omega_1\mathrm{(t)}  \widetilde{\vartheta}_1+\omega_2\mathrm{(t)}  \widetilde{\vartheta}_2+\omega_3\mathrm{(t)}\widetilde{\vartheta}_{\mathrm{3}} \right ).
\end{equation}

With this (\ref{37}), The MGE is written as:
\begin{align}\label{39}
\begin{split}
& \dt{\hat{\vartheta}}_1=\tau \hspace{0.1cm}\omega_1\mathrm{(t)}\left ( \mathrm{g}(t)-\left ( \mathrm{\omega_1(t)}\hat{\vartheta}_1+\mathrm{\omega_2(t)}\hat{\vartheta}_2+\mathrm{\omega_3(t)}\hat{\vartheta}_3 \right ) \right )\\
&\dt{\hat{\vartheta}}_2=\tau \hspace{0.1cm}\omega_2\mathrm{(t)}\left ( \mathrm{g}(t)-\left ( \mathrm{\omega_1(t)}\hat{\vartheta}_1+\mathrm{\omega_2(t)}\hat{\vartheta}_2+\mathrm{\omega_3(t)}\hat{\vartheta}_3 \right ) \right )\\
&\dt{\hat{\vartheta}}_3=\delta_{\mathrm{u}}-\dt{\widetilde{\vartheta}}_{\mathrm{3}}=(2\tau\omega_3+\tau\omega_2-2\mu\tau\omega_1)\left (\mathrm{g}-\mathrm{\omega_1(t)}\hat{\vartheta}_1-\mathrm{\omega_2(t)}\hat{\vartheta}_2-\mathrm{\omega_3(t)}\hat{\vartheta}_3  \right ).
    \end{split}
\end{align}

When the parametric error  $\widetilde{\vartheta}=\vartheta-\hat{\vartheta}$ converges to zero, the accurate parameter estimates $\hat{\vartheta}$ of $\vartheta$ using the  dynamics $\dt{\widetilde{\vartheta}}=-\dt{\hat{\vartheta}}$ is easily obtained. Thus, the final parametric error dynamics of MGE is re-written as:
\begin{align}\label{40}
    \begin{split}
   & \dt{\widetilde{\vartheta}}_1=-\tau\omega_1\left ( \hspace{0.1cm}\omega_1\mathrm{(t)}  \widetilde{\vartheta}_1+\omega_2\mathrm{(t)}  \widetilde{\vartheta}_2+\omega_3\mathrm{(t)}  \widetilde{\vartheta}_3 \right )\\
 &\dt{\widetilde{\vartheta}}_2=-\tau\omega_2\left ( \hspace{0.1cm}\omega_1\mathrm{(t)}  \widetilde{\vartheta}_1+\omega_2\mathrm{(t)}  \widetilde{\vartheta}_2+\omega_3\mathrm{(t)}  \widetilde{\vartheta}_3  \right )\\
 & \dt{\widetilde{\vartheta}}_3=-(2\tau\omega_3+\tau\omega_2-2\mu\tau\omega_1)\left ( \hspace{0.1cm}\omega_1\mathrm{(t)}  \widetilde{\vartheta}_1+\omega_2\mathrm{(t)}  \widetilde{\vartheta}_2+\omega_3\mathrm{(t)}\widetilde{\vartheta}_{\mathrm{3}} \right ).
    \end{split}
\end{align}

\begin{figure}
    \centering
    \includegraphics[width=\linewidth]{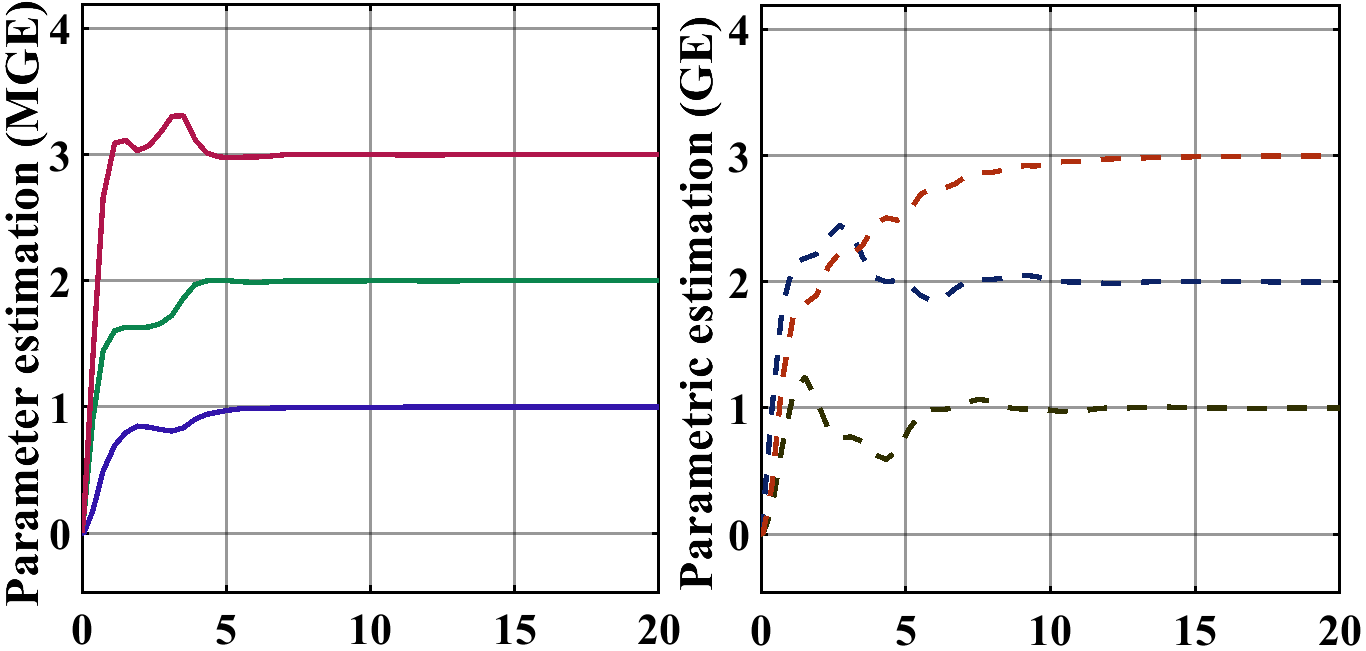}
    \caption{Time Evolution of parameter estimates and parametric error for (i) MGE:  $\mu=0.55$ and $\tau=1$ (ii) GE: $\tau=1$. Dotted lines indicate the estimation via GE and Solid lines show the estimation via MGE.}
    \label{fig3}
\end{figure}

\begin{example}\label{example3}
    Consider a PE regressor 
    \begin{equation}
        \omega(\mathrm{t})=\mathrm{col(sin(t), cos(t), sin(2t))}
    \end{equation}
     \cite{ortega2020new} with $\vartheta=\mathrm{col(1, 2, 3)}$. Based on the above error dynamics (\ref{40}), the time evolution of parametric error $\widetilde{\vartheta}(\mathrm{t})$ with the transient performance  of the proposed P\&I-based MGE  (\ref{39}) is shown in Fig. \ref{fig3}. 
\end{example}

\begin{remark}
    The proposed P\&I-based MGE provides improved results in terms of transient response and convergence speed over the conventional GE for the regressor with different cases. 
\end{remark}

\subsection{The proposed P\&I based Modified Gradient Estimator with Memory Regressor Extension (MGE with MRE)}
In GE, selecting a learning rate is complex and difficult. It has to be as low as it can be. Due to instability, a higher learning rate might lead to erroneous parameter estimations. In this proposed P\&I-based MGE, the suitable combination of the learning rate $\tau$ and $\alpha$ are required for accurate parameter estimation. In order to overcome the issues associated with the recent parameter estimation techniques and the proposed P\&I-based MGE, the proposed MGE is extended with the regressor extension procedure (Memory Regressor Extension) in this  section. An alternative approach to improve parametric convergence is proposed by Kreisselmeier in \cite{kreiselmeir}. In accordance with this approach, a single proper asymptotically stable minimum-phase transfer function i.e., first-order filter $\mathbb{L}(\mathrm{s})=\mathrm{\frac{1}{s+1}}$ is introduced in \cite{kreiselmeir}.

The underlying structure of any system dynamics can be made a problem of online estimation of unknown parameter vector appearing linearly with regressor vector as:
\begin{equation}\label{k1}
\mathrm{g}=\mathrm{\omega^T(t)}\hspace{0.075cm}\vartheta.
\end{equation}
Multiplying the above equation (\ref{k1}) by $\mathrm{\omega(t)}$ to get the matrix form as:
\begin{equation}\label{k2}
\mathrm{\omega(t)} \mathrm{g}=\mathrm{\omega(t)}\mathrm{\omega^T(t)}\hspace{0.075cm}\vartheta.
\end{equation}
This extended output (\ref{k2}) get multiplied by the simplest first-order filter $\mathbb{L}(\mathrm{s})=\mathrm{\frac{1}{s+1}}$ to get
\begin{equation}\label{k3}
    \underset{\mathrm{\mathbb{L}(s)(\omega(t)g)}}{\underbrace{\mathbb{G}}}=\underset{\mathrm{\mathbb{L}(s)(\omega(t)\omega^T(t))}}{\underbrace{\Omega} }\vartheta.
\end{equation}
Here the extended regressor $\Omega$ is given by the positive semidefinite matrix $\Omega=\mathrm{\mathbb{L}(s)(\omega(t)\omega^T(t))}$ and $\mathbb{G}=\mathrm{\mathbb{L}(s)(\omega(t)g)}$. 
The new extended LRE (\ref{k3}) motivates the following identification  algorithm
\begin{equation}\label{k4}
    \dot{\hat{\vartheta}}=\tau\left ( \mathbb{G}-\Omega \hat{\vartheta} \right )
\end{equation}
and the parametric error model
\begin{equation}\label{k5}
    \dot{\widetilde{\vartheta}}=-\tau\Omega \widetilde{\vartheta}.
\end{equation}
The main distinguishing feature of Kreisselmeier's algorithm is in use of the past value of a matrix $\omega(t)\omega^T(t)$ provided by the simplest first-order filter $\mathbb{L}(\mathrm{s})=\mathrm{\frac{1}{s+1}}$ to improve transient performance on $\hat{\vartheta}$. The main contribution of this research is to add  a virtual control law $\mathbb{\delta_{\mathrm{u}}}$ in the last equation i.e., $\dt{\widetilde{\vartheta}}_{\mathrm{q}}$. The recently developed P\&I approach \cite{ShadabArxiv} for control and stabilization is applied to get the $\mathbb{\delta_{\mathrm{u}}}$. This modifies the (\ref{k5}) to
\begin{equation}\label{k6}
    \dot{\widetilde{\vartheta}}=-\tau\Omega \widetilde{\vartheta}+\mathrm{E(q)}\mathbb{\delta_{\mathrm{u}}}
\end{equation}
with $\mathrm{E(q)}=\begin{bmatrix}
0\\ 
0\\ 
\vdots \\ 
1
\end{bmatrix}$. Here, $\mathrm{q-1}$ elements in $\mathrm{E(q)}$ are zero and only $\mathrm{q}$ element of $\mathrm{E(q)}$ is one. This addition of $\mathrm{E(q)}\mathbb{\delta_{\mathrm{u}}}$ in $ \dot{\widetilde{\vartheta}}$ provides the improved results over MRE in terms of the transient response and parameter convergence.

\textbf{The proposed P\&I-based MGE with MRE for $\mathrm{q}=2$:} The LRE for $\mathrm{q}=2$ is written as:
\begin{equation}\label{k7}
\mathrm{g}=\mathrm{\omega_1(t)}\vartheta_1+\mathrm{\omega_2(t)}\vartheta_2.
\end{equation}
Multiplying the above equation (\ref{k7}) by $\mathrm{\omega(t)}=\begin{bmatrix}
\mathrm{\omega_1(t)}\\ 
\mathrm{\omega_2(t)}
\end{bmatrix}$ to get the matrix form as:
\begin{equation}\label{k8}
\begin{bmatrix}
\mathrm{\omega_1} \mathrm{g}\\ 
\mathrm{\omega_2} \mathrm{g}
\end{bmatrix}=\begin{bmatrix}
\mathrm{\omega_1}\mathrm{\omega_1} & \mathrm{\omega_1}\mathrm{\omega_2} \\ 
 \mathrm{\omega_2}\mathrm{\omega_1}& \mathrm{\omega_2}\mathrm{\omega_2}
\end{bmatrix}\begin{bmatrix}
\vartheta_1\\
\vartheta_2
\end{bmatrix}
\end{equation}
This extended output (\ref{k8}) get multiplied by the simplest first-order filter $\mathbb{L}(\mathrm{s})=\mathrm{\frac{1}{s+1}}$ to get
\begin{equation}\label{k9}
   \underset{\mathrm{G}}{\underbrace{\begin{bmatrix}
\mathbb{G}_1\\ 
\mathbb{G}_2
\end{bmatrix}}}=\underset{\Omega}{{\underbrace{\begin{bmatrix}
\Omega_{11} & \Omega_{12} \\ 
\Omega_{21}& \Omega_{22}
\end{bmatrix}}}}\underset{\vartheta}{\underbrace{\begin{bmatrix}
\vartheta_1\\
\vartheta_2
\end{bmatrix}}}
\end{equation}
with $\mathbb{G}_1=\mathbb{L}(\mathrm{s})(\mathrm{\omega_1} \mathrm{g})$, $\mathbb{G}_2=\mathbb{L}(\mathrm{s})(\mathrm{\omega_2} \mathrm{g})$, $\Omega_{11}=\mathbb{L}(\mathrm{s})(\mathrm{\omega_1}\mathrm{\omega_1})$, $\Omega_{12}=\mathbb{L}(\mathrm{s})(\mathrm{\omega_1}\mathrm{\omega_2})$, $\Omega_{21}=\mathbb{L}(\mathrm{s})(\mathrm{\omega_2}\mathrm{\omega_1})$, and $\Omega_{22}=\mathbb{L}(\mathrm{s})(\mathrm{\omega_2}\mathrm{\omega_2})$.
The parametric error equation (\ref{k6}) for (\ref{k9}) is written as:
\begin{align}\label{k09}
    \begin{split}
        &\dot{\widetilde{\vartheta}}_1=-\tau\Omega_{11} \widetilde{\vartheta}_1-\tau\Omega_{12} \widetilde{\vartheta}_2\Rightarrow-\tau\left ( \mathbb{G}_1-\Omega_{11}\hat{\vartheta}_1-\Omega_{12}\hat{\vartheta}_2 \right )\\ 
&\dot{\widetilde{\vartheta}}_2=-\tau\Omega_{21} \widetilde{\vartheta}_1-\tau\Omega_{22} \widetilde{\vartheta}_2+\delta_{\mathrm{u}}\Rightarrow-\tau\left (\mathbb{G}_2-\Omega_{21}\hat{\vartheta}_1-\Omega_{22}\hat{\vartheta}_2 \right )+\delta_{\mathrm{u}}.
    \end{split}
\end{align}

The recently developed P\&I approach \cite{ShadabArxiv} for control and stabilization is applied to get the $\mathbb{\delta_{\mathrm{u}}}$. The very first step is the construction of the implicit manifold.
Therefore, the target dynamics $\dt{\eta}=-\tau\hspace{0.1cm}\Omega_{11}\mathrm{(t)}  \eta -\tau\mu \Omega_{12}\mathrm{(t)}  \eta$ with $\widetilde{\vartheta}_1=\eta$ is selected. This helps in defining the implicit manifold 
\begin{equation}\label{k10}
    \widetilde{\vartheta}_2-\mu \widetilde{\vartheta}_1=0.
\end{equation}
The application of four step procedure $(S_1-S_4)$ of the P\&I approach with $\alpha=0$ provides the virtual control law
\begin{equation}\label{k11}
\delta_{\mathrm{u}}=\tau\Omega_{21}\widetilde{\vartheta}_1+\tau\Omega_{22}\widetilde{\vartheta}_2-\tau\mu\Omega_{11}\widetilde{\vartheta}_1-\tau\mu\Omega_{12}\widetilde{\vartheta}_2=\tau\left ( \mathbb{G}_2-\Omega_{21}\hat{\vartheta}_1-\Omega_{22}\hat{\vartheta}_2 \right )-\tau\mu\left ( \mathbb{G}_1-\Omega_{11}\hat{\vartheta}_1-\Omega_{12}\hat{\vartheta}_2 \right ).
\end{equation}
The last equation of (\ref{k09}) can be written as:
\begin{equation}\label{gh}
    \dot{\widetilde{\vartheta}}_2=-\dot{\hat{\vartheta}}_2+\delta_{\mathrm{u}}\Rightarrow \dot{\hat{\vartheta}}_2=\delta_{\mathrm{u}}-\dot{\widetilde{\vartheta}}_2.
\end{equation}
Thus, the proposed PI\&I-based MGE with MRE
\begin{align}\label{k011}
    \begin{split}
        &\dot{\hat{\vartheta}}_1=\tau\left ( \mathbb{G}_1-\Omega_{11}\hat{\vartheta}_1-\Omega_{12}\hat{\vartheta}_2 \right )\\
&\dot{\hat{\vartheta}}_2=2\tau\left ( \mathbb{G}_2-\Omega_{21}\hat{\vartheta}_1-\Omega_{22}\hat{\vartheta}_2 \right )-\tau\mu\left ( \mathbb{G}_1-\Omega_{11}\hat{\vartheta}_1-\Omega_{12}\hat{\vartheta}_2 \right )
    \end{split}
\end{align}
is written from (\ref{gh}) and (\ref{k09}). The parametric error model of the proposed PI\&I-based MGE with MRE is written as:
\begin{align}\label{rg}
    \begin{split}
        &\dot{\widetilde{\vartheta}}_1=-\tau\Omega_{11} \widetilde{\vartheta}_1-\tau\Omega_{12} \widetilde{\vartheta}_2\\ 
&\dot{\widetilde{\vartheta}}_2=-2\tau\Omega_{21} \widetilde{\vartheta}_1-2\tau\Omega_{22} \widetilde{\vartheta}_2-\tau \mu \Omega_{11}\widetilde{\vartheta}_1-\tau \mu \Omega_{12}\widetilde{\vartheta}_2.
    \end{split}
\end{align}
\begin{example}
    Consider a not PE regressor 
    \begin{equation}
        \omega(\mathrm{t})=\mathrm{col}(1, \mathrm{((sin(t)+cos(t))/((1+t)^{0.5}))-(sin(t)/((2(1+t)^{1.5})))})
    \end{equation}
     \cite{ortega2020new} with $\vartheta=\mathrm{col(-2, 2)}$.
     \begin{figure}
    \centering
    \includegraphics[width=\linewidth]{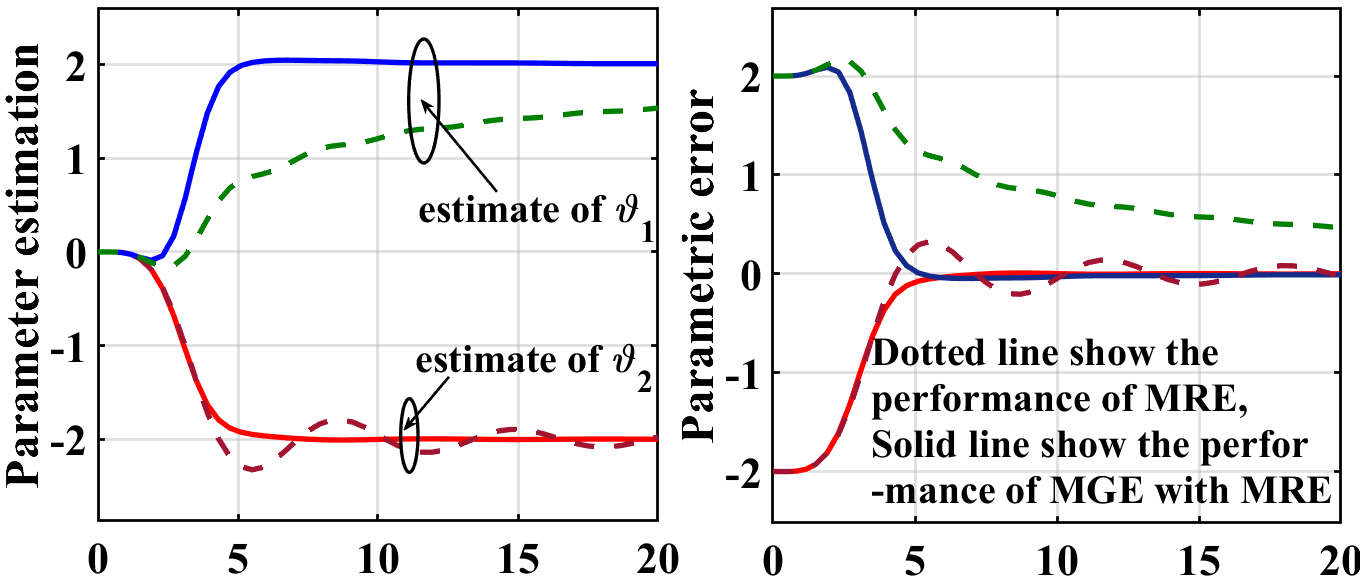}
    \caption{Time Evolution of parameter estimates and parametric error for (i) MGE:  $\mu=0.75$ and $\tau=1$ (ii) GE: $\tau=1$. Dotted lines indicate the estimation via MRE and Solid lines show the estimation via MGE with MRE.}
    \label{fig4}
\end{figure}

Based on the above error dynamics (\ref{rg}), the time evolution of parametric error $\widetilde{\vartheta}(\mathrm{t})$ with the transient performance  of the proposed P\&I-based MGE with MRE  (\ref{k011}) is shown in Fig. \ref{fig4}. 
\end{example}

\begin{example}
Consider a not PE regressor 
    \begin{equation}
        \omega(\mathrm{t})=\mathrm{col}(1, \mathrm{exp(-0.25t)})
    \end{equation}
     with $\vartheta=\mathrm{col(-2, 2)}$.
      \begin{figure}
    \centering
    \includegraphics[width=\linewidth]{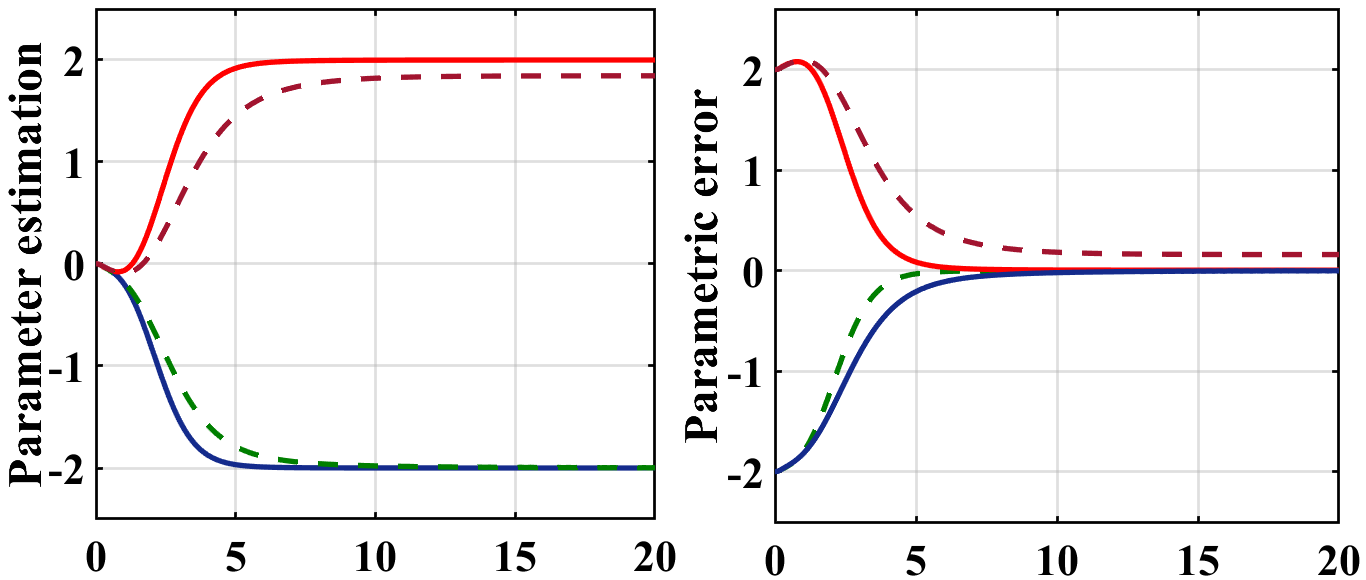}
    \caption{Time Evolution of parameter estimates and parametric error for (i) MGE:  $\mu=0.75$ and $\tau=50$ (ii) GE: $\tau=50$. Dotted lines indicate the estimation via MRE and Solid lines show the estimation via MGE with MRE.}
    \label{fig5}
\end{figure}
Based on the above error dynamics (\ref{rg}), the time evolution of parametric error $\widetilde{\vartheta}(\mathrm{t})$ with the transient performance  of the proposed P\&I-based MGE with MRE  (\ref{k011}) is shown in Fig. \ref{fig4}. The parameter error convergence to zero is guaranteed with a higher learning rate in DREM. For a similar regressor, the I-DREM \cite{glushchenko2021drem, glushchenko2021robust}ensures convergence at 60-sec \cite{glushchenko2021drem}. Based on the comparison, It is seen that the performance of the proposed PI\&I-based MGE with MRE is superior in terms of transient response and parameter convergence.
\end{example}

\begin{figure}
    \centering
    \includegraphics[width=\linewidth]{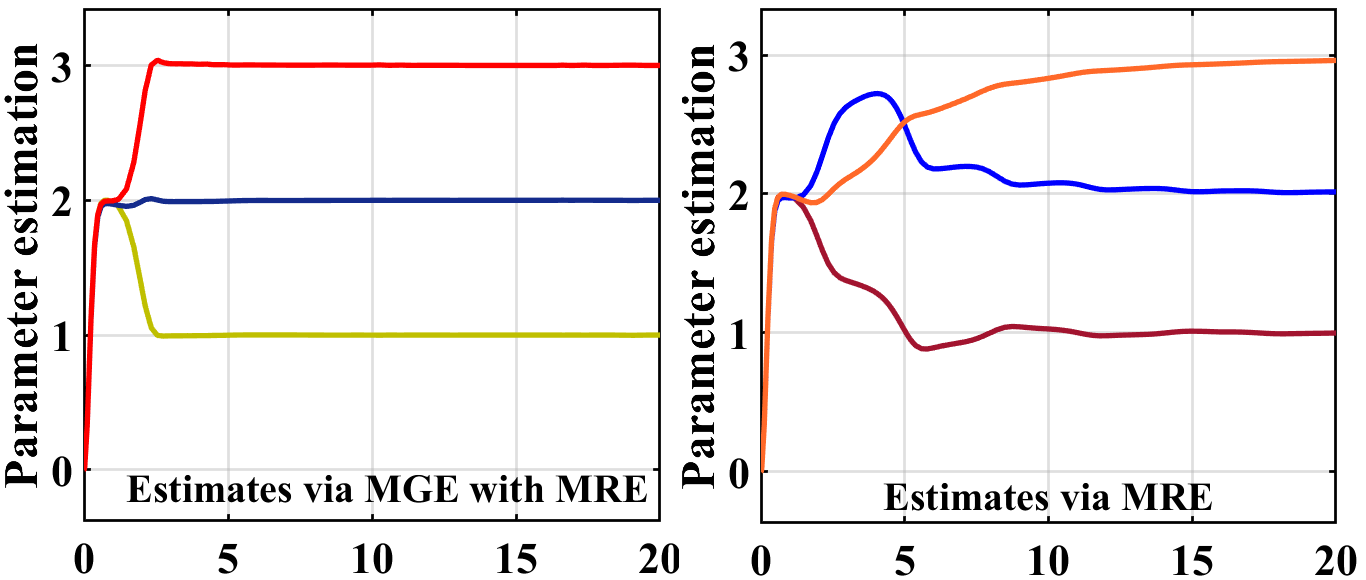}
    \caption{Time Evolution of parameter estimates and parametric error for (i) MGE with MRE:  $\mu=0.95$ and $\tau=10$ (ii) GE: $\tau=10$.}
    \label{fig6}
\end{figure}

\begin{figure}
    \centering
    \includegraphics[width=\linewidth]{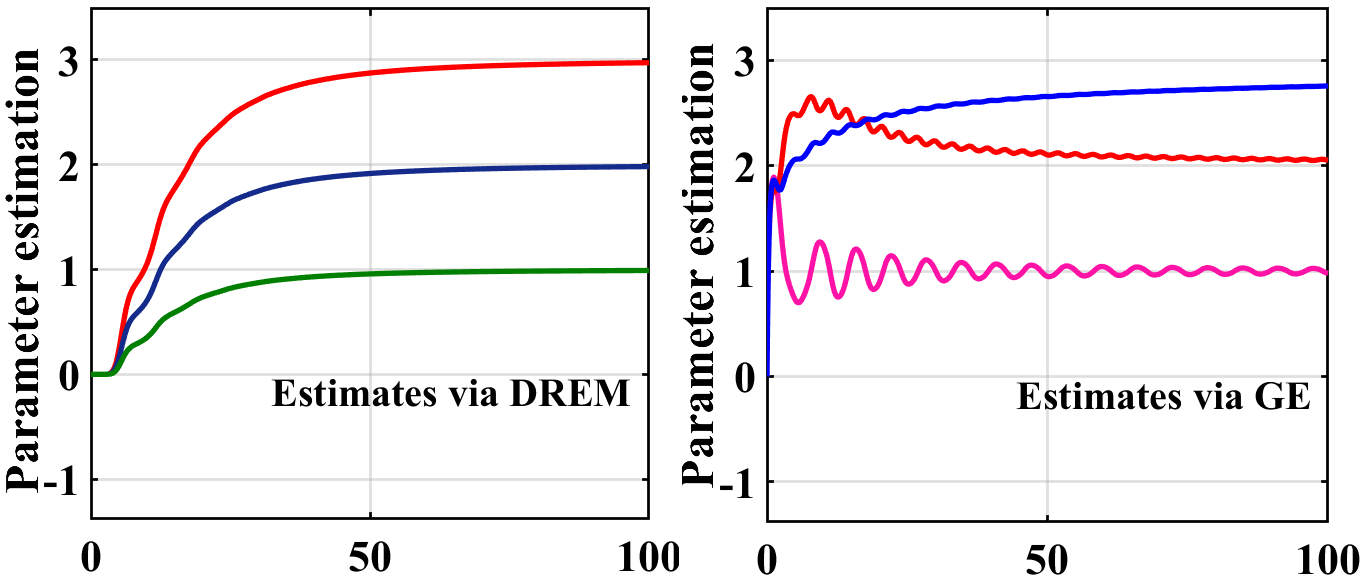}
    \caption{Time Evolution of parameter estimates via DREM and GE for $\tau=10$ and $\tau=1$ respectively.}
    \label{fig7}
\end{figure}

\textbf{The proposed P\&I-based MGE with MRE for $\mathrm{q}=3$:} The LRE for $\mathrm{q}=3$ is written as:
\begin{equation}\label{k37}
\mathrm{g}=\mathrm{\omega_1(t)}\vartheta_1+\mathrm{\omega_2(t)}\vartheta_2+\mathrm{\omega_3(t)}\vartheta_3.
\end{equation}
Multiplying the above equation (\ref{k37}) by $\mathrm{\omega(t)}=\begin{bmatrix}
\mathrm{\omega_1(t)}\\ 
\mathrm{\omega_2(t)}\\
\mathrm{\omega_3(t)}
\end{bmatrix}$ to get the matrix form as:
\begin{equation}\label{k38}
\begin{bmatrix}
\mathrm{\omega_1} \mathrm{g}\\ 
\mathrm{\omega_2} \mathrm{g}\\
\mathrm{\omega_3} \mathrm{g}
\end{bmatrix}=\begin{bmatrix}
\mathrm{\omega_1}\mathrm{\omega_1} & \mathrm{\omega_1}\mathrm{\omega_2} & \mathrm{\omega_1}\mathrm{\omega_3}\\ 
 \mathrm{\omega_2}\mathrm{\omega_1}& \mathrm{\omega_2}\mathrm{\omega_2}&
 \mathrm{\omega_2}\mathrm{\omega_3}\\
 \mathrm{\omega_3}\mathrm{\omega_1}& \mathrm{\omega_3}\mathrm{\omega_2}&
 \mathrm{\omega_3}\mathrm{\omega_3}
\end{bmatrix}\begin{bmatrix}
\vartheta_1\\
\vartheta_2\\
\vartheta_3
\end{bmatrix}
\end{equation}
This extended output (\ref{k38}) get multiplied by the simplest first-order filter $\mathbb{L}(\mathrm{s})=\mathrm{\frac{1}{s+1}}$ to get
\begin{equation}\label{k39}
   \underset{\mathrm{G}}{\underbrace{\begin{bmatrix}
\mathbb{G}_1\\ 
\mathbb{G}_2\\
\mathbb{G}_3
\end{bmatrix}}}=\underset{\Omega}{{\underbrace{\begin{bmatrix}
\Omega_{11} & \Omega_{12}&\Omega_{13} \\ 
\Omega_{21} & \Omega_{22}&\Omega_{23} \\
\Omega_{31} & \Omega_{32}&\Omega_{33}
\end{bmatrix}}}}\underset{\vartheta}{\underbrace{\begin{bmatrix}
\vartheta_1\\
\vartheta_2\\
\vartheta_3
\end{bmatrix}}}
\end{equation}
with $\mathbb{G}_1=\mathbb{L}(\mathrm{s})(\mathrm{\omega_1} \mathrm{g})$, $\mathbb{G}_2=\mathbb{L}(\mathrm{s})(\mathrm{\omega_2} \mathrm{g})$, $\mathbb{G}_3=\mathbb{L}(\mathrm{s})(\mathrm{\omega_3} \mathrm{g})$, $\Omega_{11}=\mathbb{L}(\mathrm{s})(\mathrm{\omega_1}\mathrm{\omega_1})$, $\Omega_{12}=\mathbb{L}(\mathrm{s})(\mathrm{\omega_1}\mathrm{\omega_2})$,
$\Omega_{13}=\mathbb{L}(\mathrm{s})(\mathrm{\omega_1}\mathrm{\omega_3})$,
$\Omega_{21}=\mathbb{L}(\mathrm{s})(\mathrm{\omega_2}\mathrm{\omega_1})$, $\Omega_{22}=\mathbb{L}(\mathrm{s})(\mathrm{\omega_2}\mathrm{\omega_2})$,
$\Omega_{23}=\mathbb{L}(\mathrm{s})(\mathrm{\omega_2}\mathrm{\omega_3})$,
$\Omega_{31}=\mathbb{L}(\mathrm{s})(\mathrm{\omega_3}\mathrm{\omega_1})$, $\Omega_{32}=\mathbb{L}(\mathrm{s})(\mathrm{\omega_3}\mathrm{\omega_2})$, and
$\Omega_{33}=\mathbb{L}(\mathrm{s})(\mathrm{\omega_3}\mathrm{\omega_3})$.
Defining the errors as:
\begin{align}\label{k310}
    \begin{split}
        &\epsilon_1=\mathbb{G}_1-\Omega_{11}\hat{\vartheta}_1-\Omega_{12}\hat{\vartheta}_2-\Omega_{13}\hat{\vartheta}_3=\Omega_{11}\widetilde{\vartheta}_1+\Omega_{12}\widetilde{\vartheta}_2+\Omega_{13}\widetilde{\vartheta}_3\\
&\epsilon_2=\mathbb{G}_2-\Omega_{21}\hat{\vartheta}_1-\Omega_{22}\hat{\vartheta}_2-\Omega_{23}\hat{\vartheta}_3=\Omega_{21}\widetilde{\vartheta}_1+\Omega_{22}\widetilde{\vartheta}_2+\Omega_{23}\widetilde{\vartheta}_3\\
&\epsilon_3=\mathbb{G}_3-\Omega_{31}\hat{\vartheta}_1-\Omega_{32}\hat{\vartheta}_2-\Omega_{33}\hat{\vartheta}_3 =\Omega_{31}\widetilde{\vartheta}_1+\Omega_{32}\widetilde{\vartheta}_2+\Omega_{33}\widetilde{\vartheta}_3
    \end{split}
\end{align}
The parametric error model
\begin{equation}\label{k311}
    \dot{\widetilde{\vartheta}}=-\tau\Omega \widetilde{\vartheta}.
\end{equation}
can be explicitly written as:
\begin{align}\label{k340}
    \begin{split}
        &\dot{\widetilde{\vartheta}}_1=-\tau\Omega_{11} \widetilde{\vartheta}_1-\tau\Omega_{12} \widetilde{\vartheta}_2-\tau\Omega_{13} \widetilde{\vartheta}_3=-\tau\epsilon_1\\
&\dot{\widetilde{\vartheta}}_2=-\tau\Omega_{21} \widetilde{\vartheta}_1-\tau\Omega_{22} \widetilde{\vartheta}_2-\tau\Omega_{23} \widetilde{\vartheta}_3=-\tau\epsilon_2\\
&\dot{\widetilde{\vartheta}}_3=-\tau\Omega_{31} \widetilde{\vartheta}_1-\tau\Omega_{32} \widetilde{\vartheta}_2-\tau\Omega_{33} \widetilde{\vartheta}_3=-\tau\epsilon_3
    \end{split}
\end{align}
The main contribution of this research is to add  a virtual control law $\mathbb{\delta_{\mathrm{u}}}$ in the last equation i.e., $\dt{\widetilde{\vartheta}}_{\mathrm{3}}$.
This modifies the above equation (\ref{k340}) to
\begin{align}\label{k344}
    \begin{split}
        &\dot{\widetilde{\vartheta}}_1=-\tau\Omega_{11} \widetilde{\vartheta}_1-\tau\Omega_{12} \widetilde{\vartheta}_2-\tau\Omega_{13} \widetilde{\vartheta}_3=-\tau\epsilon_1\\
&\dot{\widetilde{\vartheta}}_2=-\tau\Omega_{21} \widetilde{\vartheta}_1-\tau\Omega_{22} \widetilde{\vartheta}_2-\tau\Omega_{23} \widetilde{\vartheta}_3=-\tau\epsilon_2\\
&\dot{\widetilde{\vartheta}}_3=-\tau\Omega_{31} \widetilde{\vartheta}_1-\tau\Omega_{32} \widetilde{\vartheta}_2-\tau\Omega_{33} \widetilde{\vartheta}_3+\delta_{\mathrm{u}}=-\tau\epsilon_3+\delta_{\mathrm{u}}
    \end{split}
\end{align}
The idea used in the PI\&I approach and in the above MGE, the implicit manifold 
\begin{equation}\label{k235}
    \widetilde{\vartheta}_3+\widetilde{\vartheta}_2-2\mu \widetilde{\vartheta}_1=0.
\end{equation}
is defined in line with (\ref{35}). The application of four-step procedure $(S_1-S_4)$ of the P\&I approach with $\alpha=0$ provides the virtual control law
\begin{equation}
    \delta_{\mathrm{u}}=\tau\epsilon_3+\tau\epsilon_2-2\tau\mu\epsilon_1.
\end{equation}
The equation (\ref{k344}) can be written as:
\begin{align}\label{k345}
    \begin{split}
        &\dot{\widetilde{\vartheta}}_1=-\tau\epsilon_1=-\dot{\hat{\vartheta}}_1\\
&\dot{\widetilde{\vartheta}}_2=-\tau\epsilon_2=-\dot{\hat{\vartheta}}_2\\
&\dot{\widetilde{\vartheta}}_3=-\tau\epsilon_3+\delta_{\mathrm{u}}=-\dot{\hat{\vartheta}}_3+\delta_{\mathrm{u}}
    \end{split}
\end{align}
The last equation  $\dot{\widetilde{\vartheta}}_3=-\tau\epsilon_3+\delta_{\mathrm{u}}=-\dot{\hat{\vartheta}}_3+\delta_{\mathrm{u}}$ gives $\dot{\hat{\vartheta}}_3=\delta_{\mathrm{u}}-\dot{\widetilde{\vartheta}}_3$. Hence, the proposed PI\&I-based MGE with MRE is given by:
\begin{align}\label{k346}
    \begin{split}
        &\dot{\hat{\vartheta}}_1=\tau\epsilon_1=\tau(\mathbb{G}_1-\Omega_{11}\hat{\vartheta}_1-\Omega_{12}\hat{\vartheta}_2-\Omega_{13}\hat{\vartheta}_3)\\
&\dot{\hat{\vartheta}}_2=\tau\epsilon_2=\tau (\mathbb{G}_2-\Omega_{21}\hat{\vartheta}_1-\Omega_{22}\hat{\vartheta}_2-\Omega_{23}\hat{\vartheta}_3)\\
&\dot{\hat{\vartheta}}_3=\tau\epsilon_3+\tau\epsilon_2-2\tau\mu\epsilon_1+\tau\epsilon_3=2\tau\epsilon_3+\tau\epsilon_2-2\tau\mu\epsilon_1=2\tau(\mathbb{G}_3-\Omega_{31}\hat{\vartheta}_1-\Omega_{32}\hat{\vartheta}_2-\Omega_{33}\hat{\vartheta}_3)+\\
&\tau(\mathbb{G}_2-\Omega_{21}\hat{\vartheta}_1-\Omega_{22}\hat{\vartheta}_2-\Omega_{23}\hat{\vartheta}_3)-2\tau\mu(\mathbb{G}_1-\Omega_{11}\hat{\vartheta}_1-\Omega_{12}\hat{\vartheta}_2-\Omega_{13}\hat{\vartheta}_3)
    \end{split}
\end{align}
and corresponding parametric error model is:
\begin{align}\label{k347}
    \begin{split}
        &\dot{\widetilde{\vartheta}}_1=-\tau\epsilon_1=-\tau\Omega_{11} \widetilde{\vartheta}_1-\tau\Omega_{12} \widetilde{\vartheta}_2-\tau\Omega_{13} \widetilde{\vartheta}_3\\
&\dot{\widetilde{\vartheta}}_2=-\tau\epsilon_2=-\tau\Omega_{21} \widetilde{\vartheta}_1-\tau\Omega_{22} \widetilde{\vartheta}_2-\tau\Omega_{23} \widetilde{\vartheta}_3\\
&\dot{\widetilde{\vartheta}}_3=-(2\tau\epsilon_3+\tau\epsilon_2-2\tau\mu\epsilon_1)=-(2\tau(\Omega_{31}\widetilde{\vartheta}_1+\Omega_{32}\widetilde{\vartheta}_2+\Omega_{33}\widetilde{\vartheta}_3)+\tau(\Omega_{21}\widetilde{\vartheta}_1+\Omega_{22}\widetilde{\vartheta}_2+\Omega_{23}\widetilde{\vartheta}_3)-\\
&2\tau\mu(\Omega_{11}\widetilde{\vartheta}_1+\Omega_{12}\widetilde{\vartheta}_2+\Omega_{13}\widetilde{\vartheta}_3))
    \end{split}
\end{align}
\begin{remark}
    In the MRE, the linear stable filter is used. Roughly speaking, the application of the filter makes the second-order dynamical equation for parameter estimation. Due to this second-order dynamical system, performance enhancement is obtained. Moreover, the control perspective helps in certain modification in the final update law and provide improved results.
\end{remark}
\begin{example}
    Consider a not PE regressor 
    \begin{equation}
        \omega(\mathrm{t})=\mathrm{col}(1, \mathrm{cos(t)}, \mathrm{((sin(t)+cos(t))/((1+t)^{0.5}))-(sin(t)/((2(1+t)^{1.5})))})
    \end{equation}
 with $\vartheta=\mathrm{col(1, 2, 3)}$. Based on the above error dynamics (\ref{240}), the time evolution of parametric error $\widetilde{\vartheta}(\mathrm{t})$ with the transient performance  of the proposed P\&I-based MGE  (\ref{239}) is shown in Fig. \ref{fig6}. The DREM and GE are also applied for comparison purposes.  From Fig. \ref{fig6} and Fig. \ref{fig7}, It is easily seen that the proposed P\&I-based MGE with MRE outperforms the GE, MRE, and DREM. It improves the transient performance as well as the parameter convergence. The learning rate should be higher in the case of DREM for fast convergence. In order to overcome the issues associated with the recent approaches, the  proposed P\&I-based MGE with MRE  is proposed.
\end{example}

\section{Conclusions}\label{conclusion}
In order to get the $\mathrm{q \times q}$ square matrix, the DREM procedure requires $\mathrm{q}-1$ filters for $\mathrm{q}$-parameters. The major concern is how to carry out such an extension while maintaining the regressor's excitation level. Even if the original regressor $\omega \mathrm{(t)}$ is PE, a poor choice can compromise the convergence. The improper tuning of  the coefficient values of filters may lead to the rank-deficient matrix in DREM. The step of DREM that comprises the matrix inversion may fail. As the new regressor is based on the  determinant of an extended regression matrix, this may produce erroneous calculations and improper new extended regressor.  Moreover, the improper tuning of  the coefficient values of filters and the learning rate matrix hampers the effective implementation of DREM in certain applications. Moreover, multiple examples \cite{bobtsov2022generation} have demonstrated that consistent parameter estimate of a linear time-invariant (LTI) system with DREM is only achievable if the original regressor is PE.  To overcome the issues associated with DREM,  the GE and MRE approaches are extended by externally adding the virtual control law in the parametric error dynamics. By adding a virtual control in the final equation of the gradient dynamics, the proposed passivity and immersion-based modified gradient estimator (MGE) enhance and modify the general gradient estimator. This modification provides improved transient response and fast convergence. 
Based on certain PE and non-PE examples, a comparative analysis is carried out to show the efficacy of the proposed approaches. 

\section{Appendix 1: The P\&I Approach}\label{Appendix 2}
The  P\&I approach \cite{ShadabArxiv} is based on
\begin{itemize}
    \item an immersion of a dynamical system into a stable system using a preserving mapping and 
    \item the generation of a suitable passive output and corresponding storage function.
\end{itemize}  
This involves a particular PR structure imposed on the tangent bundle associated with the control system.

Given a single input system
\begin{align}\label{feedbackoriginalsystem}
\begin{split}
\dt{\mathrm{x}}=\mathrm{f(\mathrm{x}, \lambda)}\hspace{1.5cm}
\dt{\lambda}=\mathrm{u} 
\end{split} 
\end{align}
with $(\mathrm{x}, \lambda)\in (\mathbb{R}^{n-1},\mathbb{R})$ and without any particular structure. 
\begin{remark}
The dynamics (\ref{feedbackoriginalsystem}) is chosen  for illustration of the proposed methodology. This method can be generalized for any system which is affine in $\mathrm{u}$. 
\end{remark}
The target system, the manifold invariance condition, and the implicit manifold condition—i.e., the three essential steps of the classical I\&I approach—are merged in step $(S_1)$. 
The notion of tangent space and passivity theory is invoked to propose the P\&I approach in the following three steps $(S_2-S_4)$ to ensure the \textit{manifold attractivity}.
 
$(S_1)$ \textbf{Construction of the implicit manifold:}
Assume that there exist 
$\beta:\mathbb{R}^{\mathrm{h}}\rightarrow \mathbb{R}^{\mathrm{h}}$, $\pi: \mathbb{R}^{\mathrm{h}}\rightarrow \mathbb{R}^n$,   $\Psi:\mathbb{R}^{\mathrm{n}}\rightarrow \mathbb{R}^{\mathrm{n-h}}$
with the $\mathrm{h < n}$ such that the following statement hold.

The target dynamics $\dt{\eta}=\beta(\eta)$  with $\mathrm{x}=\eta$ is defined such that the subsystem $\dt{\mathrm{x}}=\mathrm{f(\mathrm{x}, \varphi (\mathrm{\mathrm{x}}))}$ for $\mathrm{C}^{\infty}$  mapping $\varphi(\mathrm{x}):\mathbb{R}^{\mathrm{n}}\rightarrow \mathbb{R}$ has a GES/GAS equilibrium at the origin by considering the relationship $\mathrm{\lambda}=\varphi (\mathrm{\mathrm{x}})$. This defines the implicit manifold  $\Psi(\mathrm{x}, \lambda )=\mathrm{\lambda}-\varphi (\mathrm{\mathrm{x}})=0$, the implicit manifold $\mathbb{M}=\left \{(\mathrm{x},\lambda) \in \mathbb{R}^{\mathrm{n-1}}\times\mathbb{R} | \Psi(\mathrm{x}, \lambda):=\lambda-\varphi(\mathrm{x})=0 \right \}$, and $\pi(\eta)=\mathrm{col}(\eta, \varphi(\mathrm{\eta}))$.
\begin{lemma}
The implicit manifold $\mathbb{M}$ is invariant.
\end{lemma}
\textbf{Proof:} As the relationship $\mathrm{\lambda}=\varphi (\mathrm{\mathrm{x}})$ is considered above, the manifold $\varphi (\mathrm{\mathrm{x}})-\mathrm{\lambda}=0$ is written to get 
\begin{equation}\label{gradphi}
    \triangledown \Psi(\mathrm{x}, \lambda)=\begin{bmatrix}
\frac{\partial \varphi(\mathrm{x})}{\partial \mathrm{x}} & -1
\end{bmatrix}
\end{equation}
i.e., normal to the velocity vector field  
$\begin{bmatrix}
\dt{\mathrm{x}}\\ 
\dt{\lambda}
\end{bmatrix}$. Thus, the product of (\ref{gradphi}) and the velocity vector field gives $\frac{\partial \varphi(\mathrm{x})}{\partial \mathrm{x}} \dt{\mathrm{x}}-\dt{\lambda}=0\Rightarrow \dt{\varphi}({\mathrm{x}})-\dt{\varphi}({\mathrm{x}})=0$. It means the velocity vector field is always tangent to $\mathbb{M}$. Hence, $\mathbb{M}$ is  invariant.

{$(S_2)$}  \textbf{Tangent space structure for control systems:}
Consider an n-dimensional manifold $\mathbb{M}$ with tangent bundle $\mathbb{T}_{\mathbb{M}}$, such that all $\mathrm{p}\in {\mathbb{M}}$, $\mathbb{T}_{\mathrm{p}}{\mathbb{M}}$ has the following structure
\begin{equation}
    \mathbb{T}_{\mathrm{p}}{\mathbb{M}} = \mathbb{H}_{\mathrm{p}} \oplus  \mathbb{V}_{\mathrm{p}}: \hspace{0.3cm}  \mathbb{H}_{\mathrm{p}} \cap   \mathbb{V}_{\mathrm{p}}=0
\end{equation}
where $\mathbb{H}_{\mathrm{p}}$ is the horizontal space and $\mathbb{V}_{\mathrm{p}}$ is the vertical space.  
\begin{equation}\label{TpMm}
  \mathrm{Then} \hspace{0.9cm} \mathbb{T}_{\mathrm{p}}{\mathbb{M}}= \mathbb{H}_{\mathrm{p}} \oplus  \mathbb{V}_{\mathrm{p}}=(\dt{\mathrm{x}}, 0)\oplus (0, \dt{\lambda})=(\dt{\mathrm{x}},\dt{\lambda})
\end{equation} 
is written for given system (\ref{feedbackoriginalsystem}) at any point $\mathrm{p}\in\mathbb{M}$. With the implicit manifold $\Psi (\mathrm{x} ,\lambda )$ obtained in $S_1$, the normal vector direction is given by $\triangledown \Psi (\mathrm{x} ,\lambda )$. Thus, a PR metric $\mathrm{R}$ on space $\mathbb{T}_{\mathrm{p}}{\mathbb{M}}$ can be defined as
\begin{small}
\begin{align}\label{metric}
    \begin{split}
        \mathrm{R}&=\triangledown \Psi (\mathrm{x} ,\lambda )^{\mathrm{T}}\triangledown \Psi (\mathrm{x} ,\lambda )\\&=\begin{bmatrix}
\left ( \frac{\partial \varphi}{\partial \mathrm{x}}\right )^{\mathrm{T}}\left ( \frac{\partial \varphi}{\partial \mathrm{x}}\right ) & \left ( -\frac{\partial \varphi}{\partial \mathrm{x}}\right )^{\mathrm{T}}\\ 
-\left ( \frac{\partial \varphi}{\partial \mathrm{x}}\right ) & \mathrm{I}
\end{bmatrix}=\begin{bmatrix}
\mathrm{\mathrm{m}}_{11} & \mathrm{\mathrm{m}}_{12}\\ 
 \mathrm{\mathrm{m}}_{21}&\mathrm{\mathrm{m}}_{22}
\end{bmatrix}.
    \end{split}
\end{align}
\end{small}
which is intuitively a natural choice. To proceed with the passive output, the obtained implicit manifold and a  PR  metric $\mathrm{R}$ are utilized in the  splitting of tangent space [further details in Appendix]. The off-the-manifold dynamics can now be viewed as evolving in the tangent space with metric as $\left ( \mathbb{T}_{\mathbb{M}},\mathrm{R} \right )$.
\begin{remark}
To obtain the passive output, the metric $\chi$ is replaced with semi-Riemannian metric $\mathrm{R}$ as a natural choice. (Details and proof can be found in Appendix)
\end{remark}
For $(\mathbb{M}, \mathrm{R})$, the splitting is visualized as follows:
\begin{small}
\begin{equation}
     (\dt{\mathrm{x}},\dt{\lambda})=\left (\dt{\mathrm{x}}, -\mathrm{m}_{22}^{-1}\mathrm{m}_{21} \dt{\mathrm{x}} \right )\oplus \left (0, \dt{\lambda}+\mathrm{m}_{22}^{-1}\mathrm{m}_{21} \dt{\mathrm{x}} \right ) =\widetilde{\mathbb{H}}_{\mathrm{p}} \oplus \widetilde{\mathbb{V}}_{\mathrm{p}}
\end{equation}
\end{small}

As $\dt{\lambda}$  is along the vertical direction, the passive output is chosen as a component $\dt{\lambda}+\mathrm{m}_{22}^{-1}\mathrm{m}_{21} \dt{\mathrm{x}}$ which  is in the same direction or parallel to $\dt{\lambda}$ \cite{ShadabArxiv}.  Roughly speaking, the idea is to bring the component $\dt{\lambda}+\mathrm{m}_{22}^{-1}\mathrm{m}_{21}$ of $\widetilde{\mathbb{H}}_{\mathrm{p}}$ to the component $\mathrm{m}_{22}^{-1}\mathrm{m}_{21} \dt{\mathrm{x}}$ of $\widetilde{\mathbb{V}}_{\mathrm{p}}$.
\begin{figure}[ht!]
    \centering
    \includegraphics[width=\linewidth]{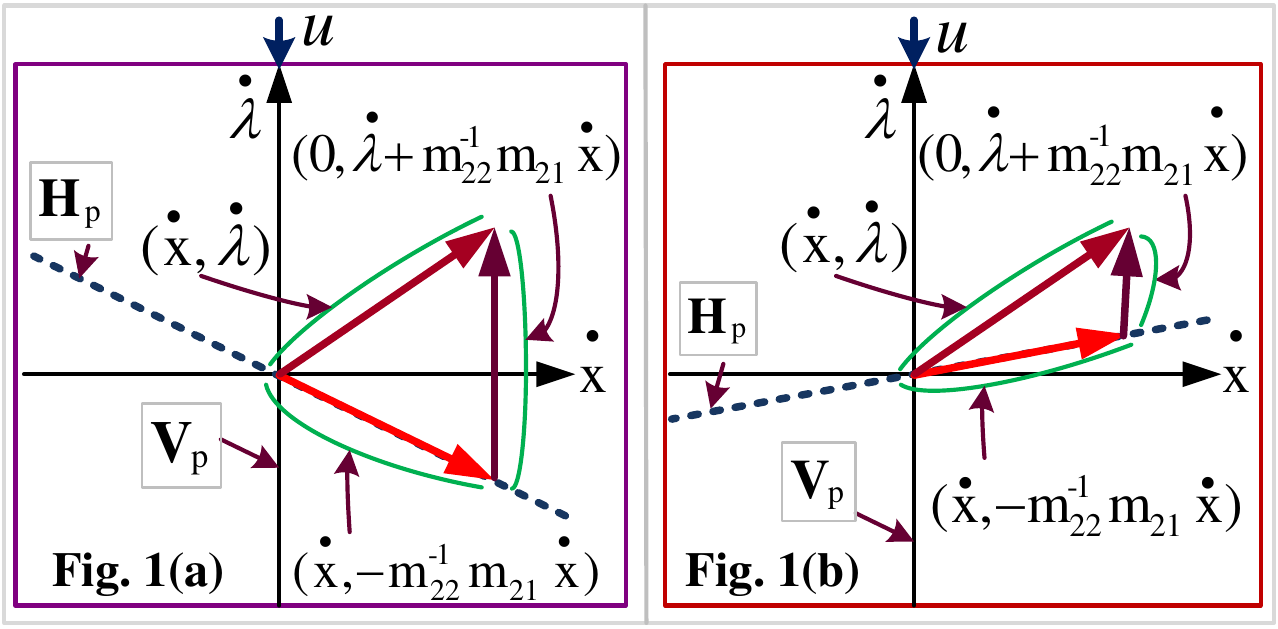}
    \caption{Geometrical interpretation: vertical vector $\mathbb{V}_{\mathrm{p}}$ is along the fiber direction and $\mathbb{H}_{\mathrm{p}} \oplus  \mathbb{V}_{\mathrm{p}}=\mathbb{T}_{\mathrm{p}}{\mathbb{M}}$. }
    \label{Geometrical interpretation}
\end{figure}
The geometrical interpretation for the splitting tangent vector is shown in Fig. \ref{Geometrical interpretation}. The different possibilities of splitting tangent vector are shown in Fig. \ref{Geometrical interpretation}(a) and Fig. \ref{Geometrical interpretation}(b). The discussion about the passive output and storage function is provided in Appendix.

\textbf{$(S_3)$  Passive output:}
The component of $\mathrm{u}$ tangent vector along $\dt{\lambda}$  is used to define the \textit{passive output} 
\begin{equation}\label{GammaPassive}
    \mathrm{y}=\mathrm{y}_1+\mathrm{y}_2=\int_{0}^{t}\dt{\lambda}\hspace{0.08cm}\mathrm{dt}
+{\int_{0}^{t}(\mathrm{m}_{22}^{-1}\mathrm{m}_{21}\dt{\mathrm{x}})\hspace{0.08cm}\mathrm{dt}}
\end{equation}
with the help of the passivity theory. Here, $\mathrm{y}_1={\int_{0}^{t}\dt{\lambda}\hspace{0.08cm}\mathrm{dt}}$ and $\mathrm{y}_2={\int_{0}^{t}(\mathrm{m}_{22}^{-1}\mathrm{m}_{21}\dt{\mathrm{x}})\hspace{0.08cm}\mathrm{dt}}$ are defined. If $\mathrm{m}_{22}^{-1}\mathrm{m}_{21}$ is a constant then $\mathrm{y}=({\lambda}+\mathrm{m}_{22}^{-1}\mathrm{m}_{21}{\mathrm{x}})$ is defined. If $\mathrm{m}_{22}^{-1}\mathrm{m}_{21}$ is a function of $\mathrm{x}$ then, it can be written as the gradient of any function $\mathrm{q(x)}$ i.e.,  $\mathrm{m}_{22}^{-1}\mathrm{m}_{21}\mathrm{(x)}=\triangledown \mathrm{q(x)}$. Then
\begin{equation}\label{Gamma}
    \mathrm{y}=\int_{0}^{t}(\dt{\lambda}+\triangledown \mathrm{q(x)} \dt{\mathrm{x}})\mathrm{dt}=(\lambda+\mathrm{q(x)})
\end{equation}
\begin{remark}
The condition  $\mathrm{m}_{22}^{-1}\mathrm{m}_{21}\mathrm{(x)}=\triangledown \mathrm{q(x)}$ is related to the condition of integrability and integrable connection in differential geometry. 
\end{remark}
In some cases, the systems will no longer have a closed-form expression. In such cases, finding the implicit manifold and passive output may be difficult or even impossible in different approaches including the P\&I method to stabilize the systems in upper triangular form.

$(S_4)$ \textbf{Storage function:}
With $\mathrm{y}$, the candidate Lyapunov function $\mathbb{S}(\mathrm{x}, \lambda)$ (i.e., storage function) is defined as
\begin{equation}\label{Storage function}
  \mathbb{S}(\mathrm{x}, \lambda)=\frac{1}{2}{\mathrm{y}}^2=\frac{1}{2}(\lambda+\mathrm{q(x)})^2.
\end{equation}
The convergence of the off-the-manifold dynamics to the implicit manifold at an exponential rate $\alpha$ is accompanied by selecting the condition
\begin{equation}\label{exponential laypunov}
    \dt{\mathbb{S}}\leq -\alpha\mathbb{S}.
\end{equation}
One can use the condition (\ref{exponential laypunov}) along with the storage function (\ref{Storage function}) and passive output (\ref{Gamma}) to get
\begin{equation}\label{exponential laypunov solution}
    (\lambda+\mathrm{q(x)})(\dt{\lambda}+\frac{\partial \mathrm{q(x)}}{\partial \mathrm{x}}\dt{\mathrm{x}})\leq -\frac{\alpha}{2}(\lambda+\mathrm{q(x)})^2
\end{equation}
\begin{equation}\label{sjksdkc}
 \Rightarrow (\dt{\lambda}+\frac{\partial \mathrm{q(x)}}{\partial \mathrm{x}}\dt{\mathrm{x}})+\frac{\alpha}{2}(\lambda+\mathrm{q(x)})=0
\end{equation}
The equation (\ref{sjksdkc}) is modified by substituting (\ref{feedbackoriginalsystem}) in terms of the final control law as
\begin{equation}\label{final control law}
  \mathrm{u}=-\frac{\alpha}{2}\lambda-\frac{\alpha}{2}\mathrm{q(x)}-\frac{\partial\mathrm{q(x)}}{\partial \mathrm{x}}\mathrm{f(\mathrm{x}, \lambda)}.
\end{equation}
The above-defined control law ensures the GAS equilibrium point of the system to zero/origin.

\begin{remark}
In the rest of this paper, the
controller of Section \ref{Appendix 2} will be referred to as the P\&I approach.
\end{remark}

\section{Appendix 2: Theorem-Proof}\label{Appendix 3}
\subsection{Splitting of Tangent Space}
Consider an n-dimensional manifold $\mathbb{M}$ with tangent bundle $\mathbb{T}_{\mathbb{M}}$, such that all $\mathrm{p}\in {\mathbb{M}}$, $\mathbb{T}_{\mathrm{p}}{\mathbb{M}}$ has the following structure
\begin{equation}
    \mathbb{T}_{\mathrm{p}}{\mathbb{M}} = \mathbb{H}_{\mathrm{p}} \oplus  \mathbb{V}_{\mathrm{p}}: \hspace{0.3cm}  \mathbb{H}_{\mathrm{p}} \cap   \mathbb{V}_{\mathrm{p}}=0.
\end{equation}
If $\mathbb{H}_{\mathrm{p}}$ horizontal space and $\mathbb{V}_{\mathrm{p}}$ the vertical space are considered then at all $\mathrm{p}\in\mathbb{M}$, $\mathbb{T}_{\mathrm{p}}{\mathbb{M}}$ is direct sum of $\mathbb{H}_{\mathrm{p}}$ and $\mathbb{V}_{\mathrm{p}}$. If $\mathbb{M}$ is coordinatized as $(\mathrm{x}, \lambda)$ with $\mathrm{x}\in \mathbb{R}^{\mathrm{k}}, \lambda \in \mathbb{R}^{\mathrm{n-k}}$, and $\mathrm{k}< \mathrm{n}$ then $\mathbb{T}_{\mathrm{p}}{\mathbb{M}}$ for any $\mathrm{p}\in\mathbb{M}$ is written as
\begin{equation}\label{TpM}
   \mathbb{T}_{\mathrm{p}}{\mathbb{M}} = \mathbb{H}_{\mathrm{p}} \oplus  \mathbb{V}_{\mathrm{p}}=(\dt{\mathrm{x}}, 0)\oplus (0, \dt{\lambda})=(\dt{\mathrm{x}},\dt{\lambda}).
\end{equation}
The interpretation of above decomposition  (\ref{TpM}) can be given in terms of $(\mathrm{M}, \chi)$, i.e., $\mathrm{M}$ is a Riemannian space with Identity $\chi$ as the Riemannian metric \cite{ShadabArxiv}. This gives
\begin{equation}
   {\left \langle \mathbb{H}_{\mathrm{p}},  \mathbb{V}_{\mathrm{p}} \right \rangle}_{\chi}= \begin{bmatrix}
\dt{\mathrm{x}} &0 
\end{bmatrix}\underset{\chi}{\underbrace{\begin{bmatrix}
\mathrm{I}^{\mathrm{k}\times \mathrm{k}} & 0\\ 
0 & \mathrm{I}^{\mathrm{n-k}\times \mathrm{n-k}}
\end{bmatrix}}}\begin{bmatrix}
0\\ 
\dt{\lambda}
\end{bmatrix}=0
\end{equation}

\begin{theorem}
For a given tangent vector $\mathrm{w}=(\dt{\mathrm{x}}, \dt{\lambda})\in \mathbb{T}_{\mathrm{p}}{\mathbb{M}}$, the orthogonality of  $\mathbb{H}_{\mathrm{p}}$ and $\mathbb{V}_{\mathrm{p}}$ is preserved under new pseudo-Riemannian metric $\mathrm{R}$.
\end{theorem}
\textbf{Proof:}   Suppose the metric $\chi$ is replaced with some other pseudo-Riemannian metric 
\begin{equation}
    \mathrm{R}=\begin{bmatrix}
\mathrm{\mathrm{m}}_{11} & \mathrm{\mathrm{m}}_{12}\\ 
 \mathrm{\mathrm{m}}_{21}&\mathrm{\mathrm{m}}_{22}
\end{bmatrix}
\end{equation}
(or (\ref{metric})) with $\mathrm{\mathrm{m}}_{12}=\mathrm{\mathrm{m}}_{21}$. Then to ensure that $\mathbb{T}_{\mathrm{p}}{\mathbb{M}} =\widetilde{\mathbb{H}}_{\mathrm{p}} \oplus \widetilde{\mathbb{V}}_{\mathrm{p}}$,  the following is the structure of $\widetilde{\mathbb{H}}_{\mathrm{p}}$ and $\widetilde{\mathbb{V}}_{\mathrm{p}}$ under $\mathrm{R}$
\begin{small}
\begin{equation}
    \mathbb{T}_{\mathrm{p}}{\mathbb{M}}\subset (\dt{\mathrm{x}},\dt{\lambda})=\left (\dt{\mathrm{x}}, -\mathrm{\mathrm{m}}_{22}^{-1}\mathrm{\mathrm{m}}_{21} \dt{\mathrm{x}} \right )\oplus \left (0, \dt{\lambda}+\mathrm{\mathrm{m}}_{22}^{-1}\mathrm{\mathrm{m}}_{21} \dt{\mathrm{x}} \right )
\end{equation}
\end{small}
becomes
\begin{small}
\begin{equation*}
    {\left \langle \mathbb{H}_{\mathrm{p}},  \mathbb{V}_{\mathrm{p}} \right \rangle}_\mathrm{R}= \begin{bmatrix}
\dt{\mathrm{x}} &-\mathrm{\mathrm{m}}_{22}^{-1}\mathrm{w}_{21}\dt{\mathrm{x}}
\end{bmatrix}\begin{bmatrix}
\mathrm{\mathrm{m}}_{11} & \mathrm{\mathrm{m}}_{12}\\ 
 \mathrm{\mathrm{m}}_{21}&\mathrm{\mathrm{m}}_{22}
\end{bmatrix}\begin{bmatrix}
0\\ 
\dt{\lambda}+\mathrm{\mathrm{m}}_{22}^{-1}\mathrm{\mathrm{m}}_{21}\dt{\mathrm{x}}
\end{bmatrix}
\end{equation*}
\end{small}
\begin{equation}
    = \mathrm{\mathrm{m}}_{21}\dt{\mathrm{x}}(\dt{\lambda}+\mathrm{\mathrm{m}}_{22}^{-1}\mathrm{\mathrm{m}}_{21}\dt{\mathrm{x}})-\mathrm{\mathrm{m}}_{21}\dt{\mathrm{x}}(\dt{\lambda}+\mathrm{\mathrm{m}}_{22}^{-1}\mathrm{\mathrm{m}}_{21}\dt{\mathrm{x}})=0
\end{equation}
\begin{remark}
The above splitting of the tangent space under a metric $\mathrm{R}$ is closely related to the theory of fiber bundles and Ehresmann connection.
\end{remark}
\begin{remark}
In differential geometry, a PR manifold is a differentiable manifold with a metric tensor that is everywhere nondegenerate. This is a generalization of a Riemannian manifold in which the requirement of positive-definiteness is relaxed.
\end{remark}
\subsection{Passive output and storage function}
\begin{lemma}
The obtained $\mathrm{y}=(\lambda+\mathrm{q(x)})$ in (\ref{Gamma}) is a passive output and associated function (\ref{Storage function}) is a storage function  with respect to  new input $\mathrm{v}$ and $\mathrm{y}$. 
\end{lemma}

\textbf{Proof:} The time-derivative of function (\ref{Storage function}) 
\begin{small}
\begin{equation}
   \dt{\mathbb{S}}= (\lambda+\mathrm{q(x)})(\dt{\lambda}+\frac{\partial \mathrm{q(x)}}{\partial \mathrm{x}}\dt{\mathrm{x}})=(\lambda+\mathrm{q(x)})(\mathrm{u}+\frac{\partial \mathrm{q(x)}}{\partial \mathrm{x}}\mathrm{f(x,\lambda)})
\end{equation}
\end{small}
is translated into
\begin{align}\label{khdsdfs}
    \begin{split}
         \dt{\mathbb{S}} 
         =\mathrm{y^T}\mathrm{u}+\mathrm{y^T}\frac{\partial \mathrm{q(x)}}{\partial \mathrm{x}}\mathrm{f(x,\lambda)}
    \end{split}
\end{align}
with $\mathrm{y}=(\lambda+\mathrm{q(x)})$. Here, $\mathrm{y}\in \mathbb{R}^{\mathrm{n-k}}$ and the dimension of $\mathrm{y}, \mathrm{u}$, and $\mathrm{v}$ is same.   By substituting the control law 
\begin{equation}\label{final control law stable}
  \mathrm{u}=-\frac{\partial\mathrm{q(x)}}{\partial \mathrm{x}}\mathrm{f(x,\lambda)}+\mathrm{v}
\end{equation} 
in (\ref{khdsdfs}), it becomes passive with respect to  new input $\mathrm{v}$ and $\mathrm{y}$ due to 
\begin{equation}
    \dt{\mathbb{S}} \leq \mathrm{y^T v }. 
\end{equation}
Therefore, the $\mathbb{S}$ is called as the storage function with $\mathrm{y}=(\lambda+\mathrm{q(x)})$ as a passive output.

\begin{remark}
Since $\mathbb{S}$ is positive definite outside the manifold $\mathbb{M}$ and  $\mathbb{S}(0,0)=0$ on the manifold, the obtained storage function (\ref{Storage function}) can be used as the candidate Lyapunov function to prove the attractivity of the manifold.
\end{remark}

\bibliography{references.bib}
\bibliographystyle{IEEEtran}

\end{document}